\documentclass[aps, prl, twocolumn, superscriptaddress, urlcolor=blue, linkcolor=blue]{revtex4-1}

\allowdisplaybreaks[1]
\pdfoutput=1
\usepackage{comment}

\newcommand{\bea}{\begin{eqnarray}}
\newcommand{\eea}{\end{eqnarray}}
\newcommand{\be}{\begin{equation}}
\newcommand{\ee}{\end{equation}}
\newcommand{\cF}{\mathcal{F}}

\newcommand{\cblue}[1]{\textcolor{black}{#1}}

\def \- {\!\smallsetminus\!}

\begin{document}

\begin{titlepage}

\title{Quantum Statistics and Spacetime Surgery 
}

%
\author{Juven Wang} \email{juven@ias.edu}
\affiliation{School of Natural Sciences, Institute for Advanced Study, Princeton, NJ 08540, USA}
\affiliation{Department of Physics, Harvard University,  Cambridge, MA 02138, USA} 
\affiliation{Center of Mathematical Sciences and Applications, Harvard University,  Cambridge, MA, 
USA} 
%
%
\author{Xiao-Gang Wen} \email{xgwen@mit.edu}
\affiliation{Perimeter Institute for Theoretical Physics, Waterloo, Ontario, N2L 2Y5, Canada} 
\affiliation{Department of Physics, Massachusetts Institute of Technology, Cambridge, MA 02139, USA}
\author{Shing-Tung Yau} \email{yau@math.harvard.edu}
\affiliation{Department of Mathematics, Harvard University, Cambridge, MA 02138, USA}
\affiliation{Center of Mathematical Sciences and Applications, Harvard University,  Cambridge, MA, 
USA} 
\affiliation{Department of Physics, Harvard University,  Cambridge, MA 02138, USA} 

\begin{abstract} 
We apply the geometric-topology surgery theory
on spacetime manifolds to study the constraints of quantum statistics data in 2+1 and 3+1 spacetime dimensions.
%
%
First, we introduce the fusion data for worldline and worldsheet operators capable creating anyon 
excitations of particles and strings,
well-defined in gapped states of matter with intrinsic topological orders.
Second, we introduce the braiding statistics data of particles and strings, such as the geometric Berry matrices for
particle-string Aharonov-Bohm and multi-loop 
adiabatic braiding process, 
encoded by submanifold linkings, in the closed spacetime 3-manifolds and 4-manifolds.
Third, we derive 
``quantum surgery'' 
constraints analogous to Verlinde formula associating 
fusion and braiding statistics data via spacetime surgery, essential for 
defining
the theory of 
topological orders, and potentially correlated to bootstrap boundary physics such as gapless modes, conformal field theories or quantum anomalies.
\end{abstract}

\maketitle
\end{titlepage}



Decades ago, the 
fractional quantum Hall effect was discovered \cite{TSG8259}.
The intrinsic relation between the topological quantum field theories (TQFT) and the topology of manifolds
was found \cite{Schwarz:1978cn, Witten:1988hf}
years after. 
The two breakthroughs  
partially 
motivated the study of topological order \cite{Wenrig} as a new state of matter in quantum many-body systems
and in condensed matter systems \cite{Wen:2012hm}. 
Topological orders are defined as the gapped states of matter with physical properties depending on 
global 
topology (such as the ground state degeneracy (GSD)), 
robust against any local perturbation and any symmetry-breaking perturbation.
Accordingly, topological orders cannot be characterized by 
the old paradigm of symmetry-breaking 
phases of matter via the Ginzburg-Landau theory \cite{GL5064,LanL58}.
The systematic studies 
of 2+1 dimensional (2+1D) topological orders enhance our understanding of the real-world plethora phases including
quantum Hall states and spin liquids \cite{BalentsSL}.
In this work, we explore the 
constraints between the 2+1D and 3+1D 
topological orders
and the geometric-topology properties of 3- 
and 4-manifolds. 
\cblue{We only focus on 2+1D / 3+1D topological orders 
with 
GSD insensitive to the system size
and
with a finite number of types of topological excitations 
creatable from 1D line and 2D surface operators.
}
\begin{figure}[!h]
\centerline{
\includegraphics[scale=0.45]{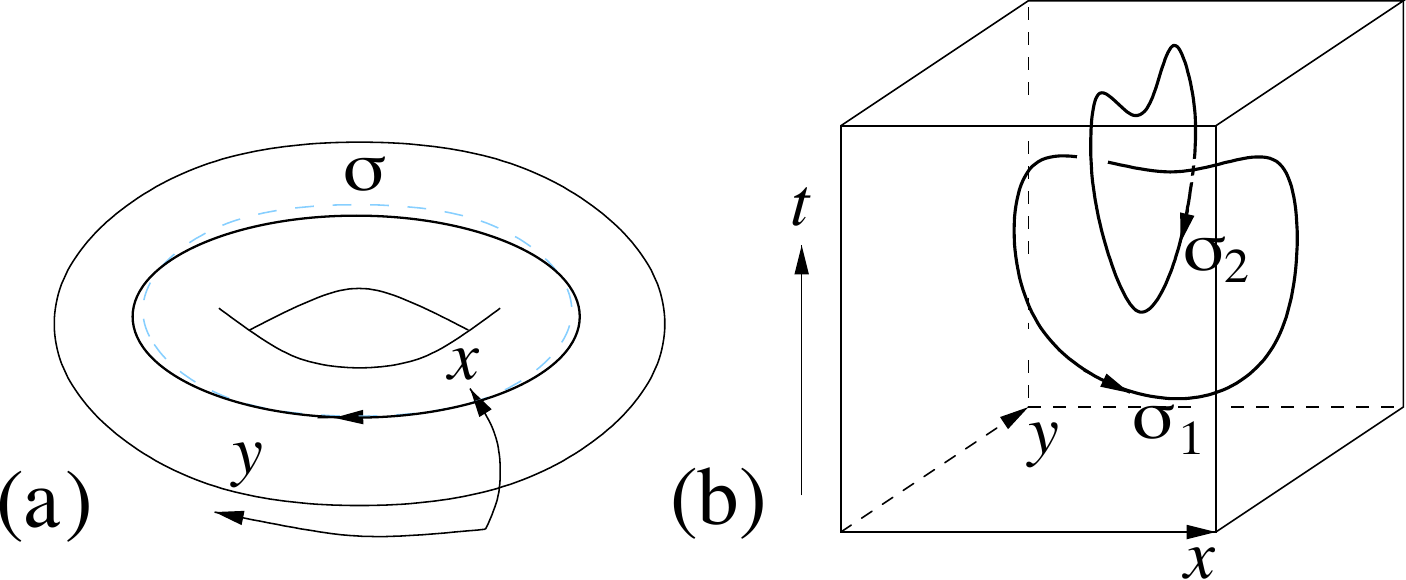}
}
\caption{
(a) A topologically-ordered ground state on 2-torus is labeled by a quasiparticle $\si$. 
(b) The quantum amplitude of two linked 
\cblue{spacetime trajectories of anyons $\si_1$ and $\si_2$}
is proportional to a complex number $\cS_{\si_1 \si_2}$. 
}
\label{lnklp}
\end{figure}

\cblue{We apply the tools of
quantum mechanics in physics and 
surgery theory in mathematics \cite{thurston1997three, gompf19994}.}
Our main results are:
(1) We provide the \emph{fusion data} for worldline and worldsheet operators creating excitations of particles (i.e. anyons \cite{Wilczek:1990ik}) and strings (i.e. anyonic strings) in topological orders.
(2) We provide the \emph{braiding statistics data} of particles and strings encoded by submanifold linking, in the 3- and 4-dimensional closed spacetime manifolds.
(3) By ``cutting and gluing'' quantum amplitudes,
we derive constraints between the fusion and braiding statistics data analogous to Verlinde formula 
\cite{Verlinde:1988sn, Moore:1988qv}  
for 
2+1 and 3+1D topological orders.


\se{{Quantum Statistics: Fusion and Braiding Data}} --
Imagine a renormalization-group-fixed-point 
topologically ordered quantum system 
on a spacetime manifold $\cM$. The 
manifold can be viewed as a long-wavelength continuous limit of certain lattice regularization of the system. 
We aim to compute the quantum amplitude from ``gluing'' one ket-state $| R\rangle$ with another bra-state $\langle L |$,
such as $\langle L | R\rangle$. 
A quantum amplitude also defines a path integral or a partition function $Z$ 
with the linking of worldlines/worldsheets on 
a $d$-manifold $\cM^d$, 
read as
\be \label{eq:qaZ}
\langle L | R\rangle=Z(\cM^d; \text{Link}[\text{worldline,worldsheet}, \dots]).
\ee
%
For example, the $|R\rangle$ state can
represent a ground state of 2-torus $T^2_{xy}$ 
if we put the system on a solid torus $D^2_{xt} \times S^1_y$ \cite{TQFTGSD} 
(see Fig.\ref{lnklp}(a) as the product space of 2-dimensional disk $D^2$ and 1-dimensional circle $S^1$). 
Note that its boundary 
is $\partial (D^2 \times S^1)=T^2$, and we can view the time $t$ along the radial direction.
We label the trivial vacuum sector without any operator insertions 
as $| 0_{D^2_{xt} \times S^1_y} \rangle$, which is trivial
respect to the measurement of any contractible line operator along $S^1_x$.
A worldline operator creates a pair of 
anyon and anti-anyon at its end points, if it forms a closed loop then it can be viewed as creating 
then annihilating a pair of anyons 
in a closed trajectory \cite{beyond_gauge}. 
Inserting a line operator $W^{S^1_y}_{\si}$ in the interior of $D^2_{xt} \times S^1_y$ gives a new state 
$W^{S^1_y}_{\si} | 0_{D^2_{xt} \times S^1_y} \rangle \equiv | \si_{D^2_{xt} \times S^1_y} \rangle$.
Here ${\si}$ denotes the anyon 
type \cite{Representation} along the oriented line, see Fig.\ref{lnklp}.
Insert all possible line operators of all $\si$ can completely span the ground state sectors for 2+1D topological order.
The gluing of
$
\< 0_{D^2_{} \times S^1_{}} |0_{D^2_{} \times S^1_{}}\>
$
computes the path integral 
$Z(S^2 \times S^1)$.
If we view the $S^1$ as a compact time,
this counts the ground state degeneracy (GSD) on 
a 2D 
spatial sphere $S^2$ without quasiparticle insertions,  
thus it is a 1-dimensional Hilbert space with $Z(S^2 \times S^1)=1$.
Similar relations hold for other dimensions, e.g. 3+1D topological orders on a $S^3$ without quasi-excitation 
yields
$
\< 0_{D^3 \times S^1} |0_{D^3 \times S^1}\>=Z(S^3 \times S^1)=1
$.



\se{2+1D Data}--
In 2+1D, we 
consider the worldline operators creating particles.
We define the \emph{fusion data} via fusing worldline operators:
\bea
\label{GW}
W^{S^1_y}_{\si_1} W^{S^1_y}_{\si_2}= \cF^\si_{\si_1\si_2} W^{S^1_y}_{\si}, \;\;\text{and }
G^\al_\si \equiv \<\al|  \si_{D^2_{xt} \times S^1_y} \rangle. \;\;
\eea
Here $G^\al_\si$ is read from the projection to a 
complete basis
$ \<\al|$.
Indeed the $W^{S^1_y}_{\si}$ 
generates all the canonical bases from $|0_{D^2_{xt} \times S^1_y}\>$.
Thus the canonical projection can be 
$
\<\al|=\<0_{D^2_{xt} \times S^1_y}| (W^{S^1_y}_{\al})^\dagger=\<0_{D^2_{xt} \times S^1_y}| (W^{S^1_y}_{\bar{\al}})
=\< \al_{D^2_{xt} \times S^1_y}| ,
$
then we have
$
G^\al_\si  
 =\< 0_{D^2_{xt} \times S^1_y} | (W^{S^1_y}_{\bar{\al}}) W^{S^1_y}_{\si} |0_{D^2_{xt} \times S^1_y}\>
=Z(S^2 \times S^1; {\bar{\al}},\si)=\del_{\al\si}$,
where a pair of particle-antiparticle $\si$ and ${\bar{\si}}$ can fuse to the vacuum.
We derive
\bea
&&\cF^\al_{\si_1\si_2} 
=\< 0_{D^2_{xt} \times S^1_y} | (W^{S^1_y}_{\bar{\al}}) W^{S^1_y}_{\si_1} W^{S^1_y}_{\si_2}|0_{D^2_{xt} \times S^1_y}\> \nonumber \\
&&=Z(S^2 \times S^1; {\bar{\al}},\si_1,\si_2) \equiv \cN^\al_{\si_1\si_2},
\eea
\cblue{where this path integral 
counts the dimension of the Hilbert space (namely the GSD or the number of channels
$\si_1$ and $\si_2$ can fuse to $\al$) on the spatial $S^2$.
This shows the fusion data $\cF^\al_{\si_1\si_2}$ is 
equivalent to the fusion rule $\cN^\al_{\si_1\si_2}$, symmetric under exchanging $\si_1$ and ${\si_2}$.} 
%

More generally we can glue the $T^2_{xy}$-boundary of $D^2_{xt} \times S^1_y$ via its mapping class group (MCG), namely $\MCG(T^2)=\SL(2,\Z)$
generated by 
\bea
\hat{\cS}=\bpm 0&-1\\1&0\epm,\;\; \; \hat{\cT}=\bpm 1&1\\0&1\epm.
\eea
The $\hat{\cS}$ identifies $(x,y) \to (-y, x)$ while $\hat{\cT}$ identifies $(x,y) \to (x+y, y)$ of $T^2_{xy}$. 
Based on Eq.(\ref{eq:qaZ}), we write down
the quantum amplitudes of the two $\SL(2,\Z)$ generators $\hat{\cS}$ and $\hat{\cT}$
projecting to degenerate ground states. 
%
We denote gluing two open-manifolds $\cM_1$ and $\cM_2$ along their boundaries $\cB$ under the MCG-transformation $\hat{\cal U}$ to a new manifold 
as $\cM_1 \cup_{\cB; \hat{\cal U}} \cM_2$ \cite{gluing}.
Then it is amusing 
to visualize
the gluing
$D^2_{} \times S^1_{} \cup_{T^2; \hat \cS} D^2_{} \times S^1_{} =S^3$
shows 
that 
the $\cS_{\bar\si_1\si_2}$ 
represents the Hopf link of two $S^1$ worldlines $\si_1$ and $\si_2$ (e.g. Fig.\ref{lnklp}(b)) in $S^3$ 
with the given orientation (in the canonical basis  
$\cS_{\bar\si_1\si_2}=\<{\si_1}_{} | \hat\cS | {\si_2}_{} \>$):
\bea
\label{Sgaga}
\cS_{\bar\si_1\si_2}  &
\equiv \<{\si_1}_{D^2_{xt}\times S^1_y} | \hat\cS | {\si_2}_{D^2_{xt}\times S^1_y} \>
=Z \bpm \includegraphics[scale=0.3]{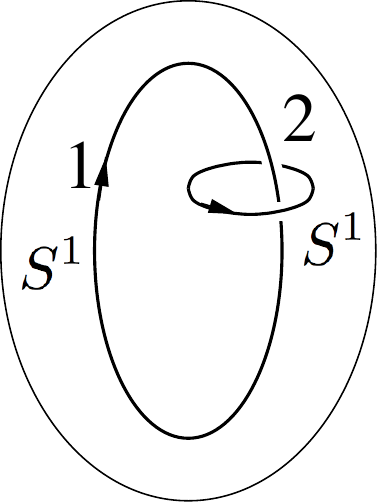} \includegraphics[scale=0.3]{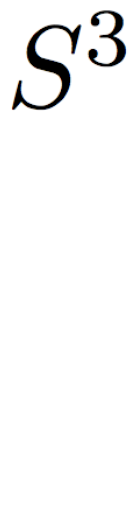} 
 \epm    
.  \;\;\;\;\;
\eea
Use the gluing $D^2_{} \times S^1_{} \cup_{T^2; \hat \cT} D^2_{} \times S^1_{}=S^2 \times S^1$,
we can derive a well known result written in the canonical bases, 
\bea
\cT_{\si_1\si_2} \equiv  \<{\si_1}_{D^2_{xt}\times S^1_y} | \hat\cT | {\si_2}_{D^2_{xt}\times S^1_y} \> 
=\del_{\si_1\si_2}\mathrm{e}^{\ii \th_{\si_2}}.\;\;\;\;\;
\eea
Its spacetime configuration is that two unlinked closed worldlines $\si_1$ and $\si_2$,
with the worldline $\si_2$ twisted by $2\pi$.
The amplitude of a twisted worldline
is given by the amplitude of untwisted worldline multiplied by  
$\mathrm{e}^{\ii \th_{\si_2}}$, where 
$\th_\si/2\pi$ is the spin of the $\si$ excitation.
It means that $\cS^\text{}_{\bar\si_1\si_2}$ measures
the \emph{mutual braiding statistics} of $\si_1$-and-$\si_2$,
while $\cT_{\si \si}$ measures the \emph{spin} and \emph{self-statistics} of $\si$.

We can introduce additional data, the Borromean rings (BR) linking between three $S^1$ circles in $S^3$, 
written as
$Z \bpm \includegraphics[scale=0.3]{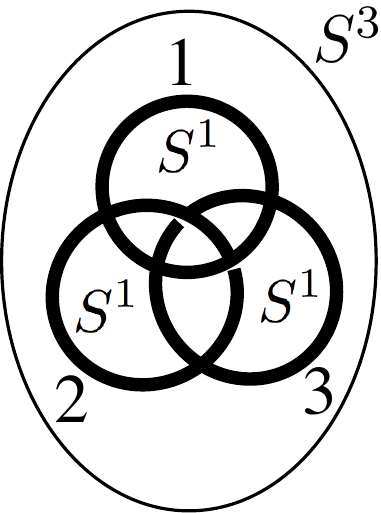} \epm$$\cblue{\equiv}Z[S^3; \text{BR}[\sigma_1,\sigma_2,\sigma_3]]$.
Although we do not know a bra-ket expression for this amplitude,
we can reduce this configuration to an easier one 
$Z[T^3_{xyt};\sigma'_{1x},\sigma'_{2y},\sigma'_{3t}]$,
a path integral 
of 3-torus 
with three orthogonal line operators each inserting along a non-contractible $S^1$ direction.
The later is a simpler expression because 
we can uniquely 
define the three line insertions exactly along the homology group generators of $T^3$, namely
$H_1(T^3,\Z)=\Z^3$. 
%
The two path integrals 
are related by three consecutive modular $\cS$ surgeries done along the $T^2$-boundary of $D^2 \times S^1$ tubular neighborhood around three $S^1$ rings \cite{S3T3}.
Namely,
$Z[T^3_{xyt};\sigma'_{1x},\sigma'_{2y},\sigma'_{3t}]=
\underset{\small{\sigma_{1},\sigma_{2},\sigma_{3}}}{\sum}
\cS_{ \sigma'_{1x} \sigma_1}
\cS_{ \sigma'_{2y} \sigma_2}
\cS_{ \sigma'_{3z} \sigma_3} Z[S^3; \text{BR}[\sigma_1,\sigma_2,\sigma_3]].$



\se{3+1D Data}--
In {3+1D}, there are intrinsic meanings of braidings of string-like excitations.
We need to consider both the worldline and the worldsheet operators which create particles and strings.
In addition to the $S^1$-worldline operator $W^{S^1}_{\si}$, 
we introduce $S^2$- and $T^2$-worldsheet operators as $V_{\mu}^{S^2}$ and $V_{\mu'}^{T^2}$ which create closed-strings (or loops) at their spatial cross sections. 
We consider the vacuum sector ground state on open 4-manifolds:
$| 0_{D^3 \times S^1} \rangle $, 
$| 0_{D^2 \times S^2} \rangle$, $| 0_{D^2 \times T^2} \rangle$ and $| 0_{S^4 \- D^2 \times T^2} \rangle$,
while their boundaries are $\partial({D^3 \times S^1})=\partial({D^2 \times S^2})=S^2 \times S^1$
and $\partial({D^2 \times T^2})=\partial({S^4 \- D^2 \times T^2})=T^3$.
Here ${\cM_1 \- \cM_2}$ means the complement space of $\cM_2$ out of $\cM_1$. 
Similar to 2+1D, we define the \emph{fusion data} $\cF^{M}$ by fusing operators: 
\bea
\label{F3+1DS1}
&&W^{S^1}_{\si_1} W^{S^1}_{\si_2}= (\cF^{S^1})^\si_{\si_1\si_2} W^{S^1}_{\si}, \\
&&V^{S^2_{}}_{\mu_1} V^{S^2_{}}_{\mu_2}=  (\cF^{S^2})_{{\mu_1}{\mu_2}}^{\mu_3} V^{S^2_{}}_{\mu_3}, \label{F3+1DS2} \\
&&V^{T^2_{}}_{\mu_1} V^{T^2_{}}_{\mu_2}=  (\cF^{T^2})_{{\mu_1}{\mu_2}}^{\mu_3} V^{T^2_{}}_{\mu_3}. \label{F3+1DT2}
\eea
Notice that we introduce additional upper indices in the fusion algebra $\cF^{M}$ to specify the topology of $M$ for the fused operators \cite{FM}.
We require normalizing worldline/sheet operators for a proper basis, 
so that the $\cF^{M}$ is also properly normalized in order for $Z(  Y^{d-1} \times S^1; \dots )$ as the GSD on
a spatial closed manifold $Y^{d-1}$ always be a positive integer.
In principle, 
we can derive the fusion rule
of excitations in any closed spacetime 4-manifold.
For instance, the fusion rule for fusing three particles on a spatial $S^3$ is
$Z(S^3 \times S^1; {\bar{\al}},\si_1,\si_2)
=\langle 0_{D^3 \times S^1} | W^{S^1}_{\bar{\al}}  W^{S^1}_{\sigma_1} W^{S^1}_{\sigma_2} | 0_{D^3 \times S^1} \rangle=
(\cF^{S^1})^\alpha_{\si_1  \sigma_2 }$.
Many more examples of fusion rules can be derived from
computing  $Z(\cM^4; {\si}, {\mu}, \dots)$ \cite{index}
by using $\cF^{M}$ and Eq.(\ref{eq:qaZ}),
here the worldline and worldsheet are submanifolds \emph{parallel not linked} 
 with each other.

If the worldline and worldsheet are linked as Eq.(\ref{eq:qaZ}), then the path integral 
encodes 
the \emph{braiding data}. Below we discuss the important braiding processes in 3+1D.
First, the Aharonov-Bohm particle-loop braiding  
can be represented as a $S^1$-worldline of particle and a $S^2$-worldsheet of loop linked in $S^4$ spacetime,
\begin{align}
\label{S2S1glue} 
{\tL}^{(S^2,S^1)}_{ \mu \sigma}
\equiv \langle 0_{D^2 \times S^2} | V_{\mu}^{S^2_{}\dagger} W^{S^1}_{\si} | 0_{D^3 \times S^1} \rangle
=Z \bpm \includegraphics[scale=0.32]{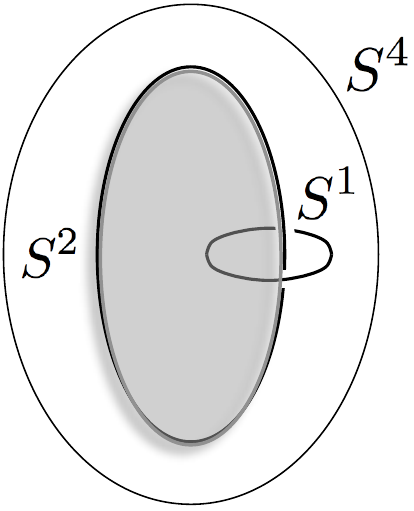} \epm,   
\end{align}
if we design the worldline and worldsheet along the generators of the first and the second homology group $H_1(D^3 \times S^1,\Z)=H_2(D^2 \times S^2,\Z)=\Z$ respectively 
via Alexander duality.
We also use the fact $ S^2_{} \times D^2_{}  \cup_{S^2 \times S^1}  D^3_{} \times S^1_{} =S^4$,
thus $\langle 0_{D^2 \times S^2 } | 0_{D^3 \times S^1} \rangle= Z(S^4)$.
Second, we can also consider 
particle-loop braiding  
as a $S^1$-worldline of particle and a $T^2$-worldsheet 
(below $T^2$ drawn as a $S^2$ with a handle) 
of loop linked in $S^4$,
\bea
\label{T2S1glue}
 \langle  0_{D^2 \times T^2}  | V_{\mu}^{T^2_{}\dagger} W^{S^1}_{\si}   | 0_{S^4 \- D^2 \times T^2} \rangle
=Z \bpm \includegraphics[scale=0.35]{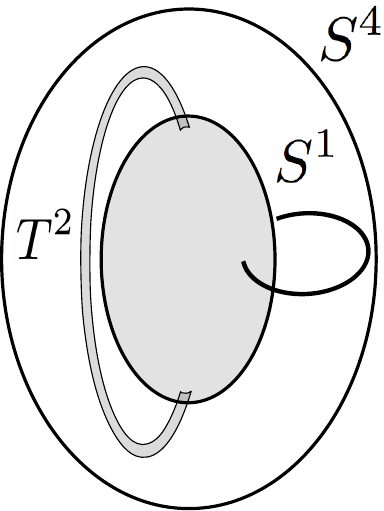} \epm,\;\;\;\;\;
\eea
if we design the worldline and worldsheet along the generators of $H_1({S^4 \- D^2 \times T^2},\Z)=H_2(D^2 \times T^2,\Z)=\Z$ respectively. 
Compare Eqs.(\ref{S2S1glue}) and (\ref{T2S1glue}),
 the loop excitation of $S^2$-worldsheet 
 is shrinkable \cite{nocharge},
while the loop of $T^2$-worldsheet needs not to be shrinkable. 

Third, 
we can 
represent 
a three-loop braiding process \cite{Wang:2014xba, Jiang:2014ksa, Moradi:2014cfa, Wang:2014oya,Jian:2014vfa, Bi:2014vaa}
as three $T^2$-worldsheets
 \emph{triple-linking} \cite{carter2004surfaces} in the spacetime $S^4$ (as the first figure in Eq.(\ref{T2T2T2glue})).
We find that
\bea \label{T2T2T2glue}
&&
{\tL^{\text{Tri}}_{{\mu_3}, {\mu_2}, {\mu_1}}} 
\equiv
\langle 0_{S^4 \- {D^2_{wx} \times T^2_{yz} }}| V^{T^2_{zx} \dagger}_{\mu_3} V^{T^2_{xy} \dagger}_{\mu_2} V^{T^2_{yz}}_{\mu_1} | 0_{{D^2_{wx} \times T^2_{yz} }} \rangle \nonumber\\
&&=Z \bpm \includegraphics[scale=0.35]{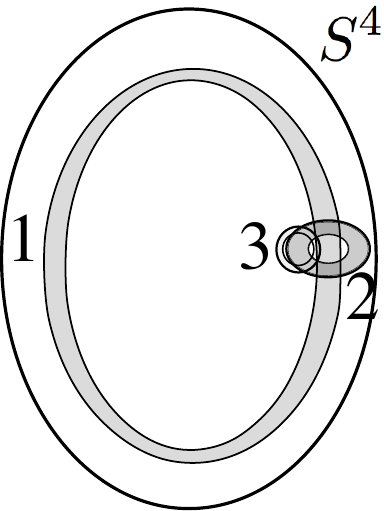} \epm
=Z \bpm \includegraphics[scale=0.35]{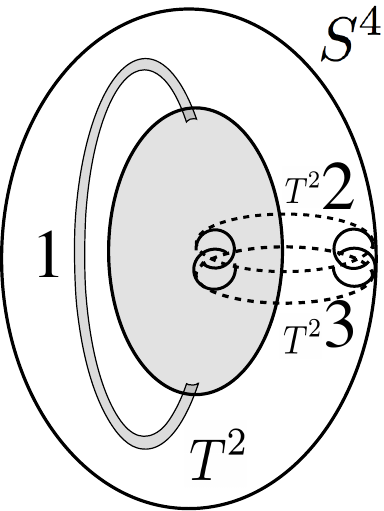} \epm,
\eea
where we design the worldsheets $V^{T^2_{yz}}_{\mu_1}$ along the generator of homology group $H_2(D^2_{wx} \times T^2_{yz},\Z)=\Z$
while we design  $V^{T^2_{xy} \dagger}_{\mu_2}$ and $V^{T^2_{zx} \dagger}_{\mu_3}$  along the two generators of $H_2({S^4\- {D^2_{wx} \times T^2_{yz}}},\Z)=\Z^2$ respectively. 
\cblue{We find that Eq.(\ref{T2T2T2glue}) 
is also equivalent to the 
spun surgery construction 
of a Hopf link (denoted as $\mu_2$ and $\mu_3$) 
linked by a third $T^2$-torus (denoted as $\mu_1$) \cite{Jian:2014vfa, Bi:2014vaa}. 
Namely, we can view the above figure as 
a Hopf link of two loops 
spinning along the dotted path of a $S^1$ circle, which
 becomes a pair of $T^2$-worldsheets $\mu_2$ and $\mu_3$.
Additionally the $T^2$-worldsheet $\mu_1$ 
(drawn in gray as a $S^2$ added a thin handle), 
together with $\mu_2$ and $\mu_3$, the three worldsheets have a 
triple-linking topological invariance \cite{carter2004surfaces}.
}
%

Fourth, the four-loop braiding process, 
where three loops dancing in the Borromean ring trajectory while linked by a fourth loop \cite{PhysRevB.91.165119}, 
can characterize certain 3+1D non-Abelian topological orders \cite{Wang:2014oya}. 
We find it is also the spun surgery 
construction of Borromean rings of three loops linked by a fourth torus in the spacetime picture,
and 
its path integral 
$Z[S^4; \text{Link[Spun[BR}[\mu_4 ,\mu_3, \mu_2]],\mu_1]]$ can be transformed:
\bea \label{ZSpinBRS4}
&& 
 Z \bpm \includegraphics[scale=0.37]{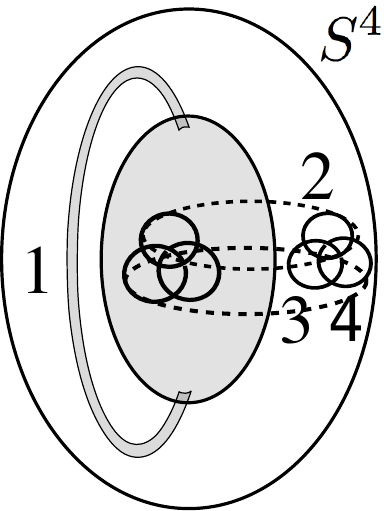} \epm
  \xrightarrow{\;\text{surgery}\;}
Z[T^4 \# S^2  \times S^2; \mu_4',\mu_3',\mu_2',\mu_1' ] \nonumber \\
&&=\langle 0_{T^4 \# S^2 \times S^2 \- {D^2_{wx} \times T^2_{yz}}} |  V^{T^2\dagger}_{\mu_4' } V^{T^2\dagger}_{\mu_3' }  V^{T^2\dagger}_{\mu_2' } V^{T^2_{yz}}_{\mu_1' } |0_{D^2_{wx} \times T^2_{yz}} \rangle,\;\;\;\;\;\;\;\;
\eea
where the surgery 
contains four consecutive modular $\cS$-transformations done along the $T^3$-boundary of $D^2 \times T^2$ tubular neighborhood around four $T^2$-worldsheets \cite{S4T4}.
The final spacetime manifold is $T^4 \# S^2  \times S^2$, where $\#$ stands for the connected sum.

We can glue the $T^3$-boundary of 4-submanifolds (e.g. $D^2 \times T^2$ and ${S^4 \- D^2 \times T^2}$) via $\MCG(T^3)=\SL(3,\Z)$ 
generated by  \cite{3+1DST} 
\bea
\hat{\cS}^{xyz}=\bpm 0& 0&1 \\1& 0&0 \\0& 1&0\epm, \;\;\; \hat{\cT}^{xy}=\bpm 1&1 &0\\0&1 & 0 \\0 &0 & 1\epm.
\eea
In this work, we define their representations as \cite{3+1DST} 
\bea
&& \label{eq:Sxyz}
{\cS^{xyz}_{\mu_2, \mu_1}} \equiv 
\< 0_{D^2_{xw} \times T^2_{yz}} | V^{T^2_{yz} \dagger}_{\mu_2}  \hat{\cS}^{xyz}  V^{T^2_{yz}}_{\mu_1}  | 0_{D^2_{xw} \times T^2_{yz}} \rangle, \\
&& {\cT^{xy}_{\mu_2, \mu_1}} 
\equiv \< 0_{D^2_{xw} \times T^2_{yz}} | V^{T^2_{yz} \dagger}_{\mu_2}  \hat{\cT}^{xy}  V^{T^2_{yz}}_{\mu_1}  | 0_{D^2_{xw} \times T^2_{yz}} \rangle,
\eea
while $\cS^{xyz}$ is a spun-Hopf link in $S^3 \times S^1$, and ${\cT}^{xy}$ is related to the \emph{topological spin} and \emph{self-statistics} of closed strings \cite{Wang:2014oya}.

%

\se{{Quantum surgery and general Verlinde formulas}} --
Now we like to derive a powerful identity for fixed-point path integrals 
of topological orders. 
If the path integral 
formed by disconnected manifolds
$M$ and $N$, denoted as $M\sqcup N$,
we have
$
Z(M\sqcup N)=Z(M)Z(N)
$.
Assume that
(1) we divide both $M$ and $N$ into two pieces such that
$M=M_U\cup_{B} M_D$, $N=N_U\cup_{B} N_D$,
and their cut 
topology (dashed $B$) 
is equivalent $B={\prt M_D}={\prt M_U}={\prt N_D}={\prt N_U}$,
and (2) the Hilbert space  on the spatial slice is 1-dimensional (namely the GSD=1)\cite{GSD1}, 
then we obtain
\bea
\label{ZMN}
&& Z \bpm \includegraphics[scale=0.33]{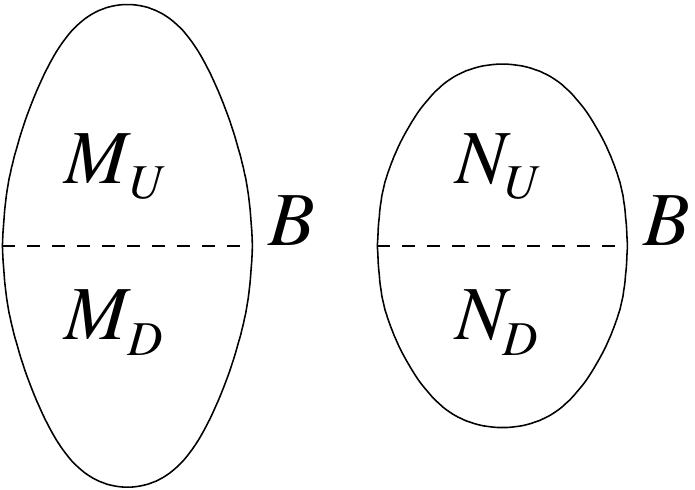} \epm 
=
 Z \bpm \includegraphics[scale=0.33]{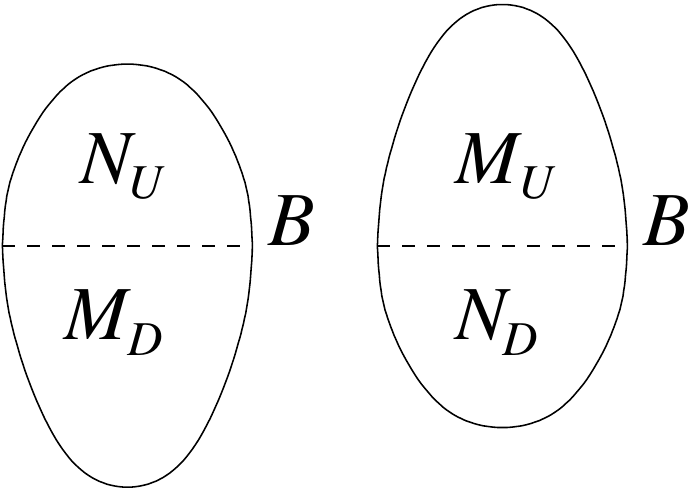} \epm 
\\
\Rightarrow
&&Z(M_U \cup_B M_D) 
Z(N_U \cup_B N_D) 
\nonumber\\
&&
 =
Z( N_U\cup_B  M_D) 
Z(M_U \cup_B N_D ). \nonumber
\eea
In 2+1D, we can derive the 
renowned Verlinde formula \cite{Witten:1988hf,  Verlinde:1988sn, Moore:1988qv}
by one version of Eq.(\ref{ZMN}): 
\bea
\label{Z2Dcut}
 Z \bpm \includegraphics[scale=0.26]{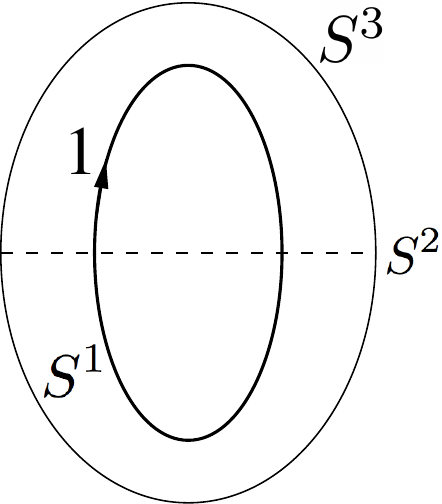} \epm 
 Z \bpm \includegraphics[scale=0.26]{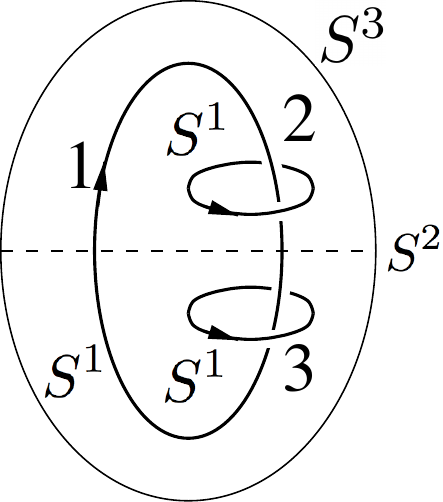} \epm 
&=&
 Z \bpm \includegraphics[scale=0.26]{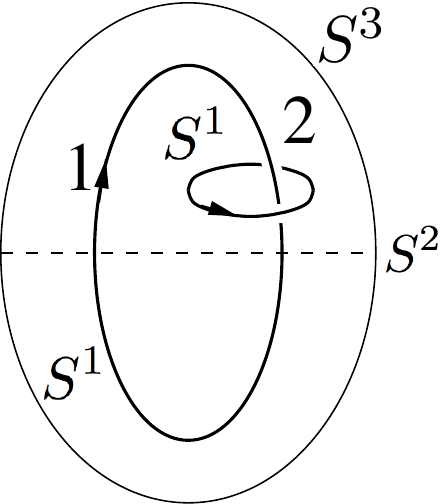} \epm 
 Z \bpm \includegraphics[scale=0.26]{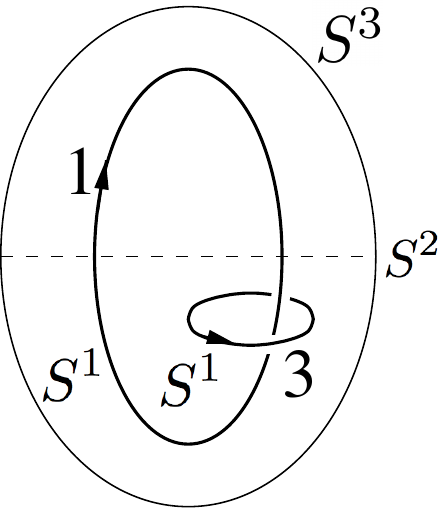} \epm \nonumber\\
 \Rightarrow \cS^\text{}_{\bar\si_10}\sum_{\si_4} \cS^\text{}_{\bar\si_1\si_4}   \cN^{\si_4}_{\si_2\si_3} 
&=&
\cS^\text{}_{\bar\si_1\si_2}
\cS^\text{}_{\bar\si_1\si_3}, 
\eea
where each spacetime manifold is $S^3$, with the line operator insertions such as an unlink and Hopf links.
Each $S^3$ is cut into two $D^3$ pieces, 
so $D^3 \cup_{S^2} D^3={S^3}$, while the boundary dashed cut is $B=S^2$.
The GSD for this spatial section $S^2$ with a pair of particle-antiparticle must be 1, so our surgery satisfies the assumptions for Eq.(\ref{ZMN}).
The second line is derived from rewriting path integrals 
in terms of our data introduced before -- the fusion rule $\cN^{\si_4}_{\si_2\si_3}$ comes from 
fusing ${\si_2\si_3}$ into ${\si_4}$ which Hopf-linked with ${\si_1}$, while Hopf links render $\cS$ matrices \cite{Verlinde}.
\cblue{
The label $0$, in $\cS^\text{}_{\bar\si_10}$ and hereafter, denotes a vacuum sector without operator insertions in a submanifold.
}

In 3+1D, the particle-string braiding in terms of $S^4$-spacetime path integral Eq.(\ref{S2S1glue}) has constraint formulas: 
\begin{widetext}
\bea
&&\label{eq:S2S1S1inS4}
 Z \bpm \includegraphics[scale=0.35]{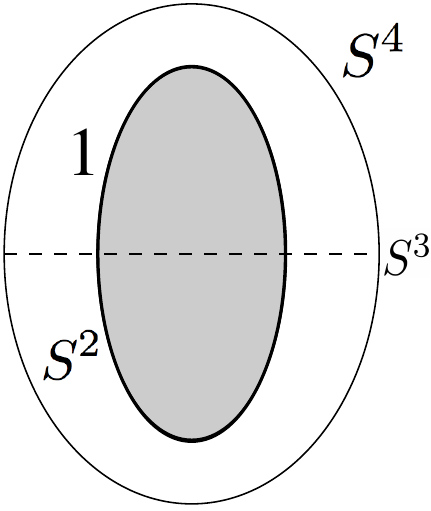}\epm  
  Z \bpm \includegraphics[scale=0.35]{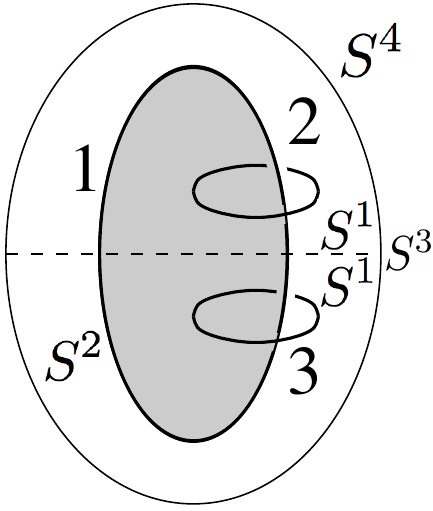} \epm 
=
 Z \bpm \includegraphics[scale=0.35]{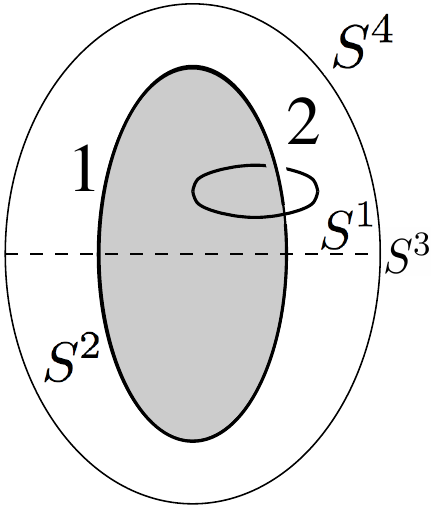} \epm 
 Z \bpm \includegraphics[scale=0.35]{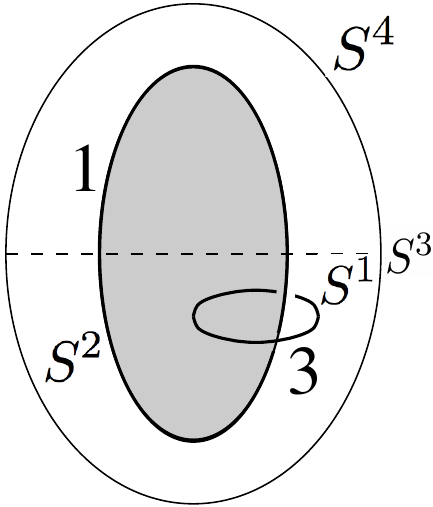} \epm 
 \Rightarrow
 {
{\tL}^\text{($S^2$,$S^1$)}_{\mu_1 0}
\sum_{\si_4}  {\tL}^\text{($S^2$,$S^1$)}_{\mu_1\si_4}  (\cF^{S^1})^{\si_4}_{\si_2\si_3}
=
{\tL}^\text{($S^2$,$S^1$)}_{\mu_1\si_2}
{\tL}^\text{($S^2$,$S^1$)}_{\mu_1\si_3}} 
 .\\
&&\label{eq:S1S2S2inS4}
  Z \bpm \includegraphics[scale=0.35]{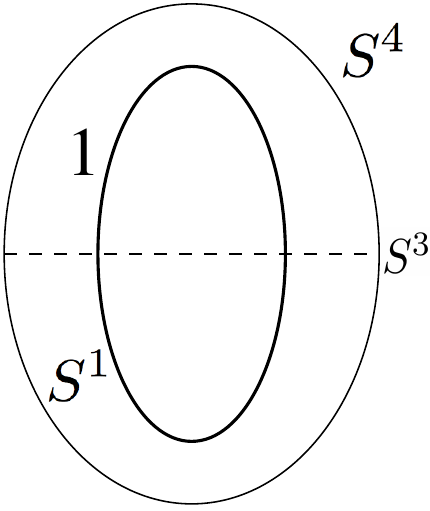} \epm 
 Z \bpm \includegraphics[scale=0.35]{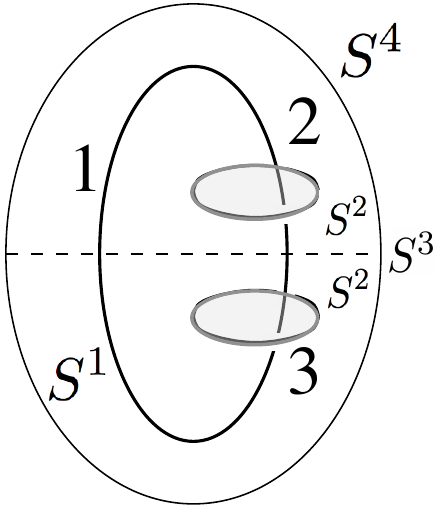} \epm 
=
 Z \bpm \includegraphics[scale=0.35]{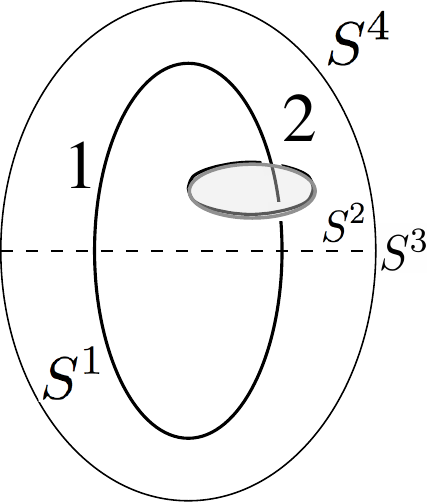} \epm 
 Z \bpm \includegraphics[scale=0.35]{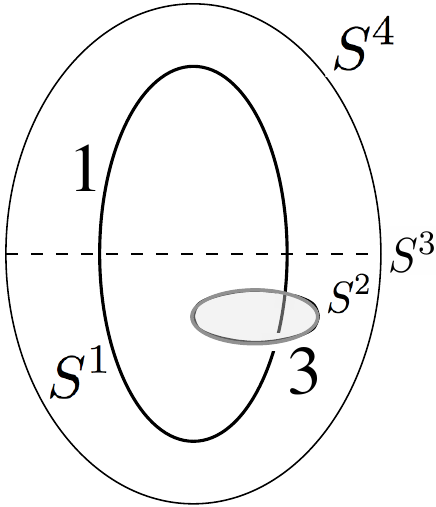} \epm 
 \Rightarrow 
  {
{\tL}^\text{($S^2$,$S^1$)}_{0 \si_1}
\sum_{\mu_4}  {\tL}^\text{($S^2$,$S^1$)}_{\mu_4 \si_1}  
(\cF^{S^2})^{\mu_4}_{\mu_2\mu_3}
=
{\tL}^\text{($S^2$,$S^1$)}_{\mu_2 \si_1}
{\tL}^\text{($S^2$,$S^1$)}_{\mu_3 \si_1}}. \;\;\;\;\;\;\;\;
 \eea
\cblue{Here the gray areas mean $S^2$-spheres.
All the data are well-defined in Eqs.(\ref{F3+1DS1}),(\ref{F3+1DS2}),(\ref{S2S1glue})}. 
Notice that Eqs.(\ref{eq:S2S1S1inS4}) and (\ref{eq:S1S2S2inS4}) are symmetric by exchanging worldsheet/worldline indices: $\mu \leftrightarrow \sigma$,
except that the fusion data is different: $\cF^{S^1}$ fuses worldlines, while $\cF^{S^2}$ fuses worldsheets.

We also derive a quantum surgery constraint formula \cite{Supple}
for the three-loop braiding in terms of $S^4$-spacetime path integral Eq.(\ref{T2T2T2glue}) via the ${\cS}^{xyz}$-surgery and its matrix representation:
%
\bea 
\label{eq:Spin[HopfLink]S4}
&&\ \ \ \
  Z \bpm \includegraphics[scale=0.45]{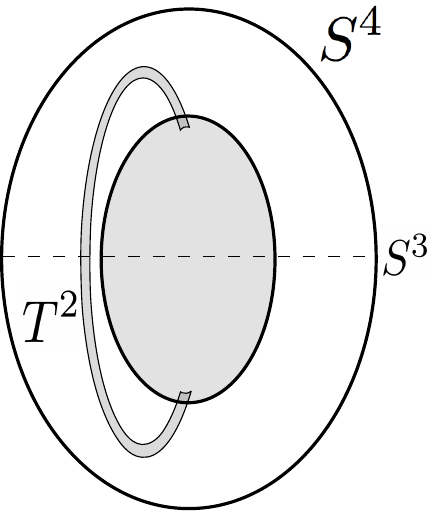} \epm  
 Z \bpm \includegraphics[scale=0.45]{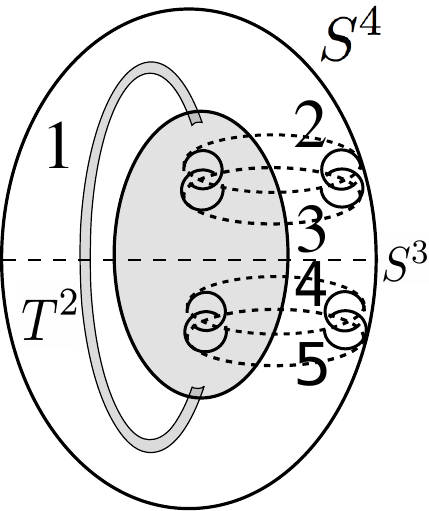} \epm  
_1=
 Z \bpm \includegraphics[scale=0.45]{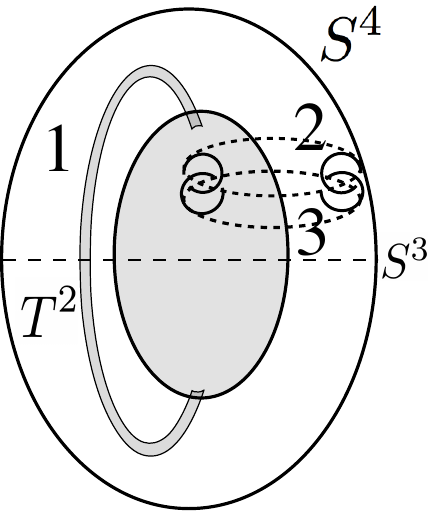} \epm 
 Z \bpm \includegraphics[scale=0.45]{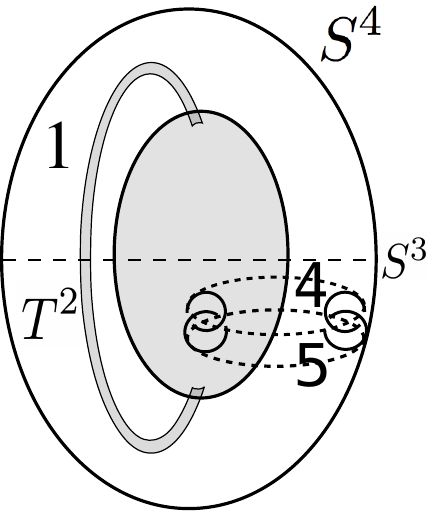} \epm \nonumber \\
&& \Rightarrow
 {{\tL^{\text{Tri}}_{0, 0, {\mu_1}}} \cdot
 \sum_{{\Gamma, \Gamma'},{\Gamma_1, \Gamma_1'}}
 (\cF^{T^2})_{{\zeta_2},{\zeta_4}}^{\Gamma}
{(\cS^{xyz})^{-1}_{\Gamma', \Gamma}}  (\cF^{T^2})_{{\mu_1} {\Gamma'}}^{\Gamma_1}  {\cS^{xyz}_{\Gamma_1', \Gamma_1}}  \; {\tL^{\text{Tri}}_{0, 0, {\Gamma_1'}}} } \nonumber
\\
&&
{=
{\sum_{{\zeta_2'}, {\eta_2}, {\eta_2'}} 
{(\cS^{xyz})^{-1}_{\zeta_2', {\zeta_2}}}  (\cF^{T^2})_{{\mu_1} {\zeta_2'}}^{\eta_2}  {\cS^{xyz}_{\eta_2', \eta_2}}  \; {\tL^{\text{Tri}}_{0, 0, {\eta_2'}}} } 
\cdot
{\sum_{{\zeta_4'}, {\eta_4}, {\eta_4'}}  
{(\cS^{xyz})^{-1}_{\zeta_4', {\zeta_4}}}  (\cF^{T^2})_{{\mu_1} {\zeta_4'}}^{\eta_4}  {\cS^{xyz}_{\eta_4', \eta_4}}  \; {\tL^{\text{Tri}}_{0, 0, {\eta_4'}}} }}, 
\eea
%
\end{widetext}
\cblue{here the ${\mu_1}$-worldsheet in gray represents a $T^2$ torus,
while ${\mu_2}$-${\mu_3}$-worldsheets and ${\mu_4}$-${\mu_5}$-worldsheets are both a pair of
two $T^2$ tori obtained by spinning the Hopf link.}
\cblue{All our data 
are well-defined in Eqs.(\ref{F3+1DT2}),(\ref{T2T2T2glue}),(\ref{eq:Sxyz}) introduced earlier.
For example, the ${\tL^{\text{Tri}}_{0, 0, {\mu_1}}}$ is defined in Eq.(\ref{T2T2T2glue})
with 0 as a vacuum without insertion, so ${\tL^{\text{Tri}}_{0, 0, {\mu_1}}}$ is a
path integral of a $T^2$ worldsheet ${\mu_1}$ in $S^4$.
}
\cblue{The index ${\zeta_2}$ is obtained from} fusing ${\mu_2}$-${\mu_3}$-worldsheets,
and \cblue{${\zeta_4}$  is obtained from} fusing ${\mu_4}$-${\mu_5}$-worldsheets.
Only ${\mu_1},{\zeta_2},{\zeta_4}$ are the fixed indices, other indices are summed over.

For all path integrals of $S^4$ in Eqs.(\ref{eq:S2S1S1inS4}), (\ref{eq:S1S2S2inS4}) and (\ref{eq:Spin[HopfLink]S4}), each $S^4$ is cut into two $D^4$ pieces, 
so $D^4 \cup_{S^3} D^4={S^4}$. \cblue{We choose all the dashed cuts for 3+1D path integral
representing $B=S^3$, while
we can view the $S^3$ as a spatial slice,
with the following excitation configurations:
A loop in Eq.(\ref{eq:S2S1S1inS4}), a pair of particle-antiparticle in Eq.(\ref{eq:S1S2S2inS4}), and a pair of loop-antiloop in Eq.(\ref{eq:Spin[HopfLink]S4}).}
Here we require a stronger criterion that all loop excitations are gapped without zero modes, 
then the GSD is 1 for all above spatial section $S^3$. 
Thus all our surgeries satisfy the assumptions for Eq.(\ref{ZMN}).

The above Verlinde-like formulas 
constrain the fusion data (e.g. $\cN$, $\cF^{S^1}$, $\cF^{S^2}$, $\cF^{T^2}$, etc.)  
and braiding data (e.g. $\cS$, $\cT$, ${\tL}^\text{($S^2$,$S^1$)}$, ${\tL^{\text{Tri}}}$, $\cS^{xyz}$, etc.). 
Moreover, we can derive constraints between the fusion data itself.
Since a $T^2$-worldsheet contains two non-contractible $S^1$-worldlines along its two homology group generators in $H_1(T^2,\Z)=\Z^2$,
the $T^2$-worldsheet operator $V^{T^2}_{\mu}$ contains the data of $S^1$-worldline operator $W^{S^1}_{\si}$. 
More explicitly, we can compute the state $W_{\sigma_1}^{S^1_y}  W_{\sigma_2}^{S^1_y} V_{\mu_2}^{T^2_{yz}} | 0_{D^2_{wx} \times T^2_{yz}} \rangle$
by fusing two $W^{S^1}_{\si}$ operators and one $V^{T^2}_{\mu}$ operator in different orders, then we obtain a consistency formula \cite{Supple}:
\bea \label{eq:FS1FT2}
\sum_{\sigma_3} (\cF^{S^1})_{{\sigma_1}{\sigma_2}}^{\sigma_3}   (\cF^{T^2})_{{\sigma_3}{\mu_2}}^{\mu_3}
=\sum_{\mu_1} (\cF^{T^2})_{{\sigma_2}{\mu_2}}^{\mu_1} (\cF^{T^2})_{{\sigma_1}{\mu_1}}^{\mu_3}. \;\;\;\;\; 
\eea
We organize 
our quantum statistics data of fusion and braiding, and some explicit examples of 
topological orders and their topological invariances in terms of our data in the Supplemental Material.


\section{Conclusion}

It is known that the quantum statistics of particles in 2+1D begets 
anyons, beyond the familiar statistics of bosons and fermions, while Verlinde formula \cite{Verlinde:1988sn}
plays a key role to dictate the consistent anyon statistics.
In this work, we derive a set of quantum surgery formulas analogous to Verlinde's constraining the fusion and braiding quantum statistics of 
anyon excitations of particle and string in 3+1D.

A further advancement of our work, comparing to the pioneer work Ref.\cite{Witten:1988hf} on 2+1D Chern-Simons gauge theory, 
is that we apply the surgery idea to generic 2+1D and 3+1D topological orders
without assuming quantum field theory (QFT) or gauge theory description. 
Although many lattice-regularized topological orders happen to have TQFT descriptions at low energy, 
we may not know which topological order derives which TQFT easily. 
Instead we simply use 
quantum amplitudes written in the bra and ket (over-)complete bases, obtained
from inserting worldline/sheet operators along the cycles of non-trivial homology group generators of a spacetime submanifold, 
to 
cut and glue to the desired path integrals. 
Consequently our approach, without the necessity of any QFT description, can be powerful to describe more generic quantum systems.
While our result is originally based on 
studying specific examples of TQFT in Dijkgraaf-Witten gauge theory \cite{Dijkgraaf:1989pz,{Supple}}, 
we formulate the 
data without using QFT. 
We have incorporated the necessary generic quantum statistic data and new constraints 
to characterize some 3+1D topological orders (including Dijkgraaf-Witten's),
we will leave the issue of their sufficiency and completeness 
for future work. 
Formally, our approach can be applied to any spacetime dimensions.

It will be interesting to study the analogous Verlinde formula constraints for 2+1D boundary states, such as highly-entangled gapless modes,
conformal field theories (CFT) and anomalies, for example through the bulk-boundary correspondence 
\cite{{Witten:1988hf}, 2015PhST164a4009R, 2015arXiv150904266C, 2015arXiv151209111W}.
The set of consistent quantum surgery formulas 
we derive may lead to an alternative effective way to \emph{bootstrap} \cite{Polyakov:1974gs,Ferrara:1973yt} 3+1D topological states of matter and 2+1D CFT.\\

\noindent
{\bf Note added}:
The formalism and some results discussed in this work have been partially reported in the first author's Ph.D. thesis \cite{JWangthesis}. 
Readers may refer to Ref.\cite{JWangthesis} for other discussions.


\section{Acknowledgements} 

We are indebted to Clifford Taubes for many generous helps on the development of this work. 
JW is grateful to 
Ronald Fintushel, 
Robert Gompf, Allen Hatcher, Shenghan Jiang, Greg Moore,
Nathan Seiberg, Ronald Stern, Andras Stipsicz, Brian Willet, Edward Witten and Yunqin Zheng for helpful comments, 
and to colleagues at 
Harvard University for discussions.
JW gratefully acknowledges the Schmidt Fellowship at IAS supported by Eric and Wendy Schmidt and the NSF Grant PHY-1314311.
This work is supported by the NSF Grant PHY-1306313, PHY-0937443, DMS-1308244, DMS-0804454,  DMS-1159412 and
Center for Mathematical Sciences and Applications at Harvard University.
This work is also supported by NSF Grant 
DMR-1506475 and NSFC 11274192, the
BMO Financial Group and the John Templeton Foundation No.\  39901.  Research at
Perimeter Institute is supported by the Government of Canada through Industry
Canada and by the Province of Ontario through the Ministry of Research.

\appendix


\begin{widetext}

\end{widetext}

\onecolumngrid

\newpage
\clearpage

\begin{center}
{\bf Supplemental Material}
\end{center}

\hfill \break
\twocolumngrid

\section{A. Summary of quantum statistics data of fusion and braiding}

\begin{table}[!h]
\begin{tabular}{l}
\hline
\hline\\[-2mm]
Quantum statistics data of fusion and braiding\\[1mm]
\hline
\hline\\[-2mm]
Data for {\bf 2+1D} topological orders:\\[1mm]
\hline\\[-2mm]
$\bullet$ Fusion data:\\[1mm]
{$\cN^{\si_3}_{\si_1 \si_2}=\cF^{\si_3}_{\si_1 \si_2}$} (fusion tensor),\\[1mm] 
$\bullet$ Braiding data:\\[1mm]
{$\cS^{xy}$, $\cT^{xy}$} (modular SL$(2,Z)$ matrices from MCG$(T^2)$),\\[1mm]
$Z[T^3_{xyt};\sigma'_{1x},\sigma'_{2y},\sigma'_{3t}]$ (or $Z[S^3; \text{BR}[\sigma_1,\sigma_2,\sigma_3]$), etc.\\[1mm]
\hline\\[-2mm]
Data for {\bf 3+1D} topological orders:\\[1mm]
\hline\\[-2mm]
$\bullet$ Fusion data:\\[1mm]
{$(\cF^{S^1})^{\si_3}_{\si_1 \si_2}$, $(\cF^{S^2})^{\mu_3}_{\mu_1 \mu_2}$, $(\cF^{T^2})^{\mu_3}_{\mu_1 \mu_2}$. } (fusion tensor)\\[1mm]
$\bullet$ Braiding data:\\[1mm]
{$\cS^{xyz}$, $\cT^{xy}$} (modular SL$(3,\Z)$ matrices from MCG$(T^3)$, \\[1mm]
including {$\cS^{xy}$})\\[1mm]
$\tL^{\text{Tri}}_{{0,0,\mu}}$ (from ${ \tL^{\text{Tri}}_{\mu_3,\mu_2,\mu_1}}$),  {$\tL^\text{Lk($S^2$,$S^1$)}_{\mu \si}$}, \\[1mm]
$Z[T^4 \# S^2  \times S^2; \mu_4',\mu_3',\mu_2',\mu_1' ]$ \\[1mm] 
(from $Z[S^4; \text{Link[Spun[BR}[\mu_4 ,\mu_3, \mu_2]],\mu_1]]$), etc. \\[1mm]
\hline
\end{tabular}
\caption{Some data for 2+1D and 3+1D topological orders encodes their quantum statistics properties, such as fusion and braiding statistics of their quasi-excitations (anyonic particles and anyonic strings).
However, the data is not complete because we do not account the degrees of freedom of their boundary modes, such as the chiral central charge $c_-=c_L-c_R$ for 2+1D  topological orders.}
\label{table:data}
\end{table}

We organize the quantum statistics data of fusion and braiding introduced in the main text in Table \ref{table:data}.
We propose using the set of data in Table \ref{table:data} to label topological orders. We also remark that
Table \ref{table:data} may not contain all sufficient data to characterize and classify all topological orders. 
What can be the missing data in Table \ref{table:data}?
Clearly, there is the chiral central charge $c_-=c_L-c_R$, the difference between the left and right central charges, missing for 2+1D  topological orders.
The $c_-$ is essential for describing 2+1D topological orders with 1+1D boundary gapless chiral edge modes.
The gapless chiral edge modes cannot be fully gapped out by adding scattering terms between different modes, because they are protected by the net chirality.
So our 2+1D data only describes \emph{2+1D non-chiral topological orders}.
Similarly, our 2+1D/3+1D data may not be able to fully classify {2+1D/3+1D topological orders} whose boundary modes are \emph{protected to be gapless}. 
We may need additional data to encode boundary degrees of freedom for their boundary modes.

In some case, some of our data may overlap with the information given by other data. For example, the 2+1D topological order data
($\cS_{xy}, \cT_{xy}$, $\cN^{\si_3}_{\si_1 \si_2}$) may contain the information of $Z[T^3_{xyt};\sigma'_{1x},\sigma'_{2y},\sigma'_{3t}]$ (or $Z[S^3; \text{BR}[\sigma_1,\sigma_2,\sigma_3]$) already,
since we know that we the former set of data may fully classify 2+1D bosonic topological orders.

Although it is possible that there are extra required data beyond what we list in Table \ref{table:data}, we find that Table \ref{table:data} is sufficient enough for
a large class of topological orders, at least for those described by Dijkgraaf-Witten twisted gauge theory \cite{Dijkgraaf:1989pz} and those gauge theories with finite Abelian gauge groups.
In the next Appendix, 
we will give some explicit examples of 2+1D and 3+1D topological orders described by Dijkgraaf-Witten theory, which can be completely characterized and classified by the data given in Table \ref{table:data}.

\begin{widetext}

\onecolumngrid
\section{B. Examples of topological orders and their topological invariances in terms of our data \label{sec:example}}

\begin{center}
\begin{table}[!h]
\noindent
\makebox[\textwidth][r]{
\begin{tabular}{ccc} 
\hline
$\begin{matrix}
\text{(i). Path-integral linking 
invariants;}\\
\text{Quantum statistic braiding data} 
\end{matrix}$ & 
$\begin{matrix}\text{(ii). Group-cohomology cocycles}\\ 
\text{distinguished by the braiding in (i)} 
\end{matrix}$ & 
$\begin{matrix}\text{(iii). TQFT actions \cblue{$\mathbf{S}$} characterized}\\ \text{by the spacetime-braiding in (i)} 
\end{matrix}$ \\
\hline\\[-2mm]
\multicolumn{3}{c}{2+1D} \\
\cline{1-3}\\[-2mm]
$\begin{matrix}
Z \bpm \includegraphics[scale=0.3]{S3ll12_uncut_l.pdf} \includegraphics[scale=0.24]{S3_top.pdf}\epm  \\   
=Z[S^3; \text{Hopf}[\sigma_1,\sigma_2]]\\
=\cS^{}_{\bar{\sigma}_1\sigma_2}
 \end{matrix}$
& $\exp \Big( \frac{2 \pi \ii {p_{IJ} }  }{N_{I} N_{J}} \; a_{I}(b_{J} +c_{J} -[b_{J} +c_{J}]) \Big) $  &  
$\begin{matrix}
\int 
\frac{ N_I}{2\pi}{B^I  \wedge \dd A^I} + { \frac{ p_{IJ}}{2 \pi}} A^I \wedge \dd A^J\\[2mm]
A^I \to A^I+ \dd g^I, \\[2mm]
N_I B^I  \to N_I B^I + \dd \eta^I.
 \end{matrix}
$  \\
\hline\\[-2mm]
$\begin{matrix}
Z \bpm \includegraphics[scale=0.35]{2+1D_Borromean_mid_123_l.pdf} \epm\\
=Z[S^3; \text{BR}[\sigma_1,\sigma_2,\sigma_3];\\
\text{Also } Z[T^3_{xyt};\sigma'_{1x},\sigma'_{2y},\sigma'_{3t}]
 \end{matrix}$      &   $ \exp \Big( \frac{2 \pi \ii p_{123}  }{N_{123}} \;  a_{1}b_{2}c_{3} \Big)$   & 
$\begin{matrix}
\int 
\frac{ N_I}{2\pi}{B^I  \wedge \dd A^I}+{{} c_{123}} A^1 \wedge A^2 \wedge  A^3\\[2mm]
A^I \to A^I+ \dd g^I, \\[2mm]
N_I B^I  \to N_I B^I + \dd \eta^I+ 2\pi {\tilde{c}}_{IJK}  A^J  g^K \\[2mm] 
- \pi {\tilde{c}}_{IJK}  g^J \dd g^K.
\end{matrix}$      \\
\hline\\[-2mm]
\multicolumn{3}{c}{3+1D} \\
\cline{1-3}\\[-2mm]
$\begin{matrix}
Z \bpm \includegraphics[scale=0.35]{Link_S2_S1_in_S4.pdf} \epm\\ 
={\tL}^{(S^2,S^1)}_{ \mu \sigma}
\end{matrix}$
& 1 & 
$
 \begin{matrix}
\int \frac{ N_I}{2\pi}{B^I  \wedge \dd A^I}\\[2mm]
A^I \to A^I+ \dd g^I, \\[2mm]
N_I B^I  \to N_I B^I + \dd \eta^I.
 \end{matrix}
$ \\
\hline\\[-2mm]
$\begin{matrix}
Z \bpm \includegraphics[scale=0.35]{3+1D_Triple_link_mid_SpinHopfLink_Large_S2_123_l.pdf} \epm \\
=Z[S^4; \text{Link[Spun[Hopf}[\mu_3,\mu_2]],\mu_1]] \\
={\tL^{\text{Tri}}_{{\mu_3}, {\mu_2}, {\mu_1}}}
\end{matrix}$ &   
${\exp \big( \frac{2 \pi \ii p_{{IJK}}^{} }{ (N_{IJ} \cdot N_K  )   }    (a_I b_J )( c_K +d_K - [c_K+d_K  ]) \big)}$   & 
$\begin{matrix}
\int 
\frac{ N_I}{2\pi}{B^I  \wedge \dd A^I} {{+}}  
\overset{}{\underset{{I,J}}{\sum}}
\frac{ N_I N_J \; p_{IJK}}{{(2 \pi)^2 } N_{IJ}}   
A^I \wedge A^J \wedge \dd A^K \\[2mm]
A^I  \to A^I + \dd g^I, \\[2mm]
 N_I B^I  \to N_I B^I  + \dd \eta^I +    \epsilon_{IJ}\frac{ N_I N_J \; p_{IJK}}{{2 \pi } N_{IJ}} d g^J \wedge A^K, \\[2mm] 
\text{here $K$ is fixed.} 
\end{matrix}$      
\\
\hline\\[-2mm]
$\begin{matrix}
Z \bpm \includegraphics[scale=0.35]{3+1D_SpinBorromean_mid_S2_1234.pdf} \epm\\
=Z[S^4; \text{Link[Spun[BR}[\mu_4 ,\mu_3, \mu_2]],\mu_1]];\\
\text{Also }  Z[T^4 \# S^2  \times S^2; \mu_4',\mu_3',\mu_2',\mu_1' ]
\end{matrix}$ & 
$\exp \big( \frac{2 \pi \ii p_{1234}}{ N_{1234} }  a_1 b_2 c_3 d_4 \big)$ & 
$\begin{matrix}
\int  \frac{ N_I}{2\pi}{B^I  \wedge \dd A^I} + {{} c_{1234}} A^1 \wedge A^2 \wedge A^3 \wedge A^4 \\[2mm]
 A^I  \to A^I + \dd g^I, \\[2mm]
 N_I B^I  \to N_I B^I + d\eta^I -\pi {\tilde{c}}_{IJKL}  A^J  A^K g^L \\[2mm]
 + \pi {\tilde{c}}_{IJKL}  A^J  g^K d g^L 
- \frac{\pi}{3} {\tilde{c}}_{IJKL}  g^J  dg^K d g^L. 
\end{matrix}
$ \\
\hline
\end{tabular}
}\hspace*{-16mm}
\caption{Examples of topological orders and their topological invariances in terms of our data in the spacetime dimension $d+1$D.
Here some explicit examples are given as Dijkgraaf-Witten twisted gauge theory \cite{Dijkgraaf:1989pz} 
with finite 
gauge group, such as $G=\Z_{N_1} \times \Z_{N_2} \times \Z_{N_3} \times \Z_{N_4} \times \dots$, 
although our quantum statistics data can be applied to more generic quantum systems without gauge or field theory description.
The first column shows the path integral form which encodes the braiding process of particles and strings in the spacetime.
In terms of spacetime picture, the path integral has nontrivial linkings of worldlines and worldsheets. The geometric Berry phases produced from
this adiabatic braiding process of particles and strings yield the measurable quantum statistics data. This data also serves as topological invariances for topological orders.
The second column shows the group-cohomology cocycle data $\omega$ as a certain partition-function solution of Dijkgraaf-Witten theory, where $\omega$ belongs to
the group-cohomology group, $\omega \in \cH^{d+1}[G,\R/\Z]=\cH^{d+1}[G,\mathrm{U}(1)]$.
The third column shows the proposed continuous low-energy field theory action form for these theories and their gauge transformations.
In 2+1D, $A$ and $B$ are 1-forms, while $g$ and $\eta$ are 0-forms.
In 3+1D, $B$ is a 2-form, $A$ and $\eta$ are 1-forms, while $g$ is a 0-form.
Here $I,J,K \in \{1,2,3,\dots\}$ belongs to the gauge subgroup indices, 
$N_{12\dots u} \equiv  \gcd(N_1,N_2, \dots, N_u)$ is defined as the greatest common divisor (gcd) of $N_1,N_2, \dots, N_u$.
Here $p_{IJ} \in \Z_{N_{IJ}}, p_{123} \in \Z_{N_{123}}, p_{IJK} \in \Z_{N_{IJK}}, p_{1234} \in \Z_{N_{1234}}$ are integer coefficients. The $c_{IJ}, c_{123}, c_{IJK}, c_{1234}$ are quantized coefficients labeling distinct topological gauge theories, where
$c_{12}=\frac{1}{(2 \pi)} \frac{N_1 N_2\;
p_{12 }}{N_{12}}$, $c_{123}=\frac{1}{(2 \pi)^2 } \frac{N_1 N_2 N_3\;
p_{123}}{N_{123}}$,  $c_{1234}=\frac{1}{(2 \pi)^3}  \frac{N_1 N_2
N_3 N_4\; p_{1234 }}{N_{1234}}$.
Be aware that we define
both $p_{IJ \dots}$ and $c_{IJ \dots}$ as constants with \emph{fixed-indices} ${I,J, \dots}$ without summing over those indices; 
while we additionally define
${\tilde{c}}_{IJ \dots} \equiv \epsilon_{IJ \dots} c_{12 \dots}$ with the $\epsilon_{IJ \dots}= \pm 1$ as an anti-symmetric Levi-Civita alternating tensor where
${I,J, \dots}$ are \emph{free indices} needed to be Einstein-summed over, but $c_{12 \dots}$ is fixed.
The lower and upper indices need to be summed-over, for example $\int \frac{ N_I}{2\pi}{B^I  \wedge \dd A^I}$ means that
$\int \overset{s}{\underset{{I=1}}{\sum}}  \frac{ N_I}{2\pi}{B^I  \wedge \dd A^I}$ where the value of $s$ depends on the total number $s$ of gauge subgroups $G=\prod_i^{s} \Z_{N_i}$. 
The quantization labelings are described and derived in \cite{{Wang:2014oya},{JuvenSPT1}}.
}
\label{table:TQFT}
\end{table}
\end{center}

In Table \ref{table:TQFT}, 
we give some explicit examples of 2+1D and 3+1D topological orders from Dijkgraaf-Witten twisted gauge theory.
We like to emphasize that our quantum-surgery Verlinde-like formulas apply to generic  2+1D and 3+1D topological orders beyond the gauge theory or field theory description.
So our formulas apply to quantum phases of matter or theories beyond the Dijkgraaf-Witten twisted gauge theory description. 
We list down these examples only because these are famous examples with a more familiar gauge theory understanding.
In terms of topological order language, Dijkgraaf-Witten theory describes the low energy physics of certain bosonic topological orders which 
can be regularized on a lattice Hamiltonian \cite{{Wang:2014oya},{Jiang:2014ksa},{Wan:2014woa}} with local bosonic degrees of freedom (without fermions).

We also clarify that what we mean by the correspondence between the items in the same row in Table \ref{table:TQFT}:
%
\begin{itemize}
\item
\cblue{(i)} 
Quantum statistic braiding data, 
\item 
\cblue{(ii)} Group cohomology cocycles 
\item 
\cblue{(iii)} 
Topological quantum field theory (TQFT). 
\end{itemize}
What we mean is that we can distinguish
the topological orders of given cocycles of (ii) with the low energy TQFT of (iii) by measuring their quantum statistic Berry phase 
under the prescribed braiding process in the path integral of (i). 
\cblue{The 
path integral of (i) is defined through the action $\mathbf{S}$ of (iii)
via
$$
Z=\int [DB_I][DA_I]\exp[ \ii \mathbf{S}].
$$
}
For example, the mutual braiding (Hopf linking) measures the $\cS$ matrix
distinguishing different types of  $\int \frac{\ii N_I}{2\pi}{B^I  \wedge d A^I} + { \frac{{\ii} p_{IJ}}{2 \pi}} A^I \wedge dA^J$ with different $p_{IJ}$ couplings;
while the Borromean ring braiding can distinguish different types of $\int \frac{\ii N_I}{2\pi}{B^I  \wedge d A^I}+{{\ii} c_{123}} A^1 \wedge A^2 \wedge  A^3$
with different $c_{123}$ couplings. However, the table does not mean that we cannot use braiding data in one row to measures the TQFT in another row.
For example, 
$\cS$ matrix can also distinguish the $\int \frac{\ii N_I}{2\pi}{B^I  \wedge d A^I}+{{\ii} c_{123}} A^1 \wedge A^2 \wedge  A^3$-type theory.
However, $Z[S^3; \text{BR}[\sigma_1,\sigma_2,\sigma_3]=Z[T^3_{xyt};\sigma'_{1x},\sigma'_{2y},\sigma'_{3t}]=1$ is trivial for 
$\int \frac{\ii N_I}{2\pi}{B^I  \wedge d A^I} + { \frac{{\ii} p_{IJ}}{2 \pi}} A^I \wedge dA^J$ with any $p_{IJ}$. Thus Borromean ring braiding cannot
measure nor distinguish the nontrivial-ness of $p_{IJ}$-type theories.

The relevant field theories are also discussed in Ref. \cite{Kapustin:2014zva, JuvenSPT1, Gaiotto:2014kfa, Gu:2015lfa, Ye:2015eba}, 
here we systematically summarize and claim
the field theories in Table  \ref{table:TQFT}  third column  describe the low energy TQFTs of Dijkgraaf-Witten theory.  

\end{widetext}
\twocolumngrid

\section{C. Derivations of some quantum surgery formulas}
In this Appendix, we derive some Verlinde-like quantum surgery formulas, which are constraints of fusion and braiding data of topological orders.
We will work out the derivations of Eqs.(\ref{Z2Dcut}),(\ref{eq:S2S1S1inS4}),(\ref{eq:S1S2S2inS4}) and then later we will derive Eq.(\ref{eq:Spin[HopfLink]S4}) step by step.
We will also derive the fusion constraint Eq.(\ref{eq:FS1FT2}) more explicitly.

First we derive a generic formula for our use of surgery. We consider a closed manifold $M$ glued by two pieces $M_U$ and $M_D$ so that
$M=M_U \cup_B M_D$ where $B=\partial M_U =  \partial M_D$.
We consider there are insertions of operators in $M_U$ and $M_D$.
We denote the generic insertions in $M_U$ as $\alpha_{M_U}$ and 
the generic insertions in $M_D$ as $\beta_{M_D}$. Here both
$\alpha_{M_U}$ and $\beta_{M_D}$ may contain both worldline and worldsheet operators.
We write the path integral 
as ${Z(M; \alpha_{M_U}, \beta_{M_D})}=\<  \alpha_{M_U} | \beta_{M_D}\>$, 
while the worldline/worldsheet may be linked or may not be linked in $M$.
Here we introduce an extra subscript $M$ in ${Z(M; \alpha_{M_U}, \beta_{M_D})}=\<  \alpha_{M_U} | \beta_{M_D}\>_M$ to specify the glued manifold is $M_U \cup_B M_D=M$. 
Now we like to do surgery by cutting out the submanifold $M_D$ out of $M$ and re-glue it back to $M_U$ via its mapping class group generator
$\hat{K} \in \MCG(B)= \MCG(\partial M_U) =   \MCG(\partial M_D)$. 
We now give some additional assumptions.\\
Assumption 1: The operator insertions in $M$ are well-separated into $M_U$ and $M_D$, so that no operator insertions cross the
boundary $B$.
Namely, at the boundary cut $B$ there are no defects of point or string excitations from the
cross-section of $\alpha_{M_U}$, $\beta_{M_D}$ or any other operators.
\\
\noindent
Assumption 2: We can generate the complete bases of degenerate ground states
fully spanning the dimension of Hilbert space for the spatial section of $B$, by inserting distinct operators (worldline/worldsheet, etc.) into $M_D$.
Namely, we insert a set of operators $\Phi$ in the interior of $| 0_{M_D}\> $ to obtain a new state $\Phi | 0_{M_D}\> \equiv | \Phi_{M_D}\>$, 
such that these states $\{\Phi | 0_{M_D}\> \}$ are orthonormal canonical bases, and
the dimension of the vector space $\dim(\{\Phi | 0_{M_D}\> \})$ equals to the ground state degeneracy ($\GSD$) of the topological order on the spatial section $B$.

If both assumptions hold, then we find a relation:
\begin{widetext}
\bea \label{}
&&{Z(M; \alpha_{M_U}, \beta_{M_D})}=\<  \alpha_{M_U} | \beta_{M_D}\>_M  =\sum_{\Phi} \langle \alpha_{M_U} | \hat{K}  \Phi | 0_{M_D}\> \< 0_{M_D}| (\hat{K}  \Phi)^\dagger |\beta_{M_D} \rangle  \nonumber \\
&&=\sum_{\Phi} \langle \alpha_{M_U} | \hat{K}  \Phi | 0_{M_D}\> \< 0_{M_D}| \Phi^\dagger \hat{K}^{-1} |\beta_{M_D} \rangle  =\sum_{\Phi} \langle \alpha_{M_U} | \hat{K}   | \Phi _{M_D}\>_{ M_U \cup_{B; \hat{K}} M_D} 
\< \Phi_{M_D}|  \hat{K}^{-1} |\beta_{M_D} \rangle_{ M_D \cup_{B; \hat{K}^{-1}} M_D}  \nonumber \\
&&=\sum_{\Phi} {Z({ M_U \cup_{B; \hat{K}} M_D}; \alpha_{M_U}, \Phi_{M_D})} \< \Phi_{M_D}|  \hat{K}^{-1} |\beta_{M_D} \rangle_{ M_D \cup_{B; \hat{K}^{-1}} M_D} 
=\sum_{\Phi} {K}^{-1}_{\Phi, \beta} \; {Z({ M_U \cup_{B; \hat{K}} M_D}; \alpha_{M_U}, \Phi_{M_D})}.
\eea
\end{widetext}
We note that in the second equality 
we write the identity matrix as $\mathbb{I}=\sum_{\Phi} ( \hat{K}  \Phi ) | 0_{M_D}\> \< 0_{M_D}| (\hat{K}  \Phi)^\dagger$.
In the third and fourth equalities 
that we have $\hat{K}^{-1}$ in the inner product $ \< \Phi_{M_D}|\hat{K}^{-1} |\beta_{M_D} \rangle$,
because $\hat{K}$ as a MCG generator acts on the spatial manifold $B$ directly.  
The evolution process from the first $\hat{K}^{-1}$ on the right and the second $\hat{K}$ on the left can be viewed as the \emph{adiabatic evolution} of
quantum states in the case of \emph{fixed-point} topological orders.
In the fifth equality 
we rewrite 
$\langle \alpha_{M_U} | \hat{K}   | \Phi _{M_D}\>_{ M_U \cup_{B; \hat{K}} M_D}={Z({ M_U \cup_{B; \hat{K}} M_D}; \alpha_{M_U}, \Phi_{M_D})}$
where $\alpha_{M_U}$ and $\Phi_{M_D}$ may or may not be linked in the new manifold ${ M_U \cup_{B; \hat{K}} M_D}$.
In the sixth equality, 
we assume that both $|\beta_{M_D} \rangle$ and $|\Phi_{M_D} \rangle$ are vectors in a canonical basis, then
we can define 
\be \label{eq:projectK}
\< \Phi_{M_D}|  \hat{K}^{-1} |\beta_{M_D} \rangle_{ M_D \cup_{B; \hat{K}^{-1}} M_D} \equiv {K}^{-1}_{\Phi, \beta}
\ee 
as a matrix element of ${K}^{-1}$, which now becomes a representation of MCG in the quasi-excitation bases of $\{|\beta_{M_D} \rangle, |\Phi_{M_D} \rangle, \dots \}$.
It is important to remember that ${K}^{-1}_{\Phi, \beta}$ is a quantum amplitude computed in the specific spacetime manifold ${ M_D \cup_{B; \hat{K}^{-1}} M_D}$.

\begin{widetext}
To summarize, so far we derive,
\be \label{eq:surger1}
\boxed{ {Z(M; \alpha_{M_U}, \beta_{M_D})}=\sum_{\Phi} {K}^{-1}_{\Phi, \beta} \; {Z({ M_U \cup_{B; \hat{K}} M_D}; \alpha_{M_U}, \Phi_{M_D})}}.  
\ee
%
%
We can also derive another formula by applying the inverse transformation,
\be \label{eq:surger2}
\boxed{ {Z({ M_U \cup_{B; \hat{K}} M_D}; \alpha_{M_U}, \Phi'_{M_D})} = \sum_{\Phi'} {K}^{}_{\beta, \Phi'} \; {Z(M; \alpha_{M_U}, \beta_{M_D})}}. 
\ee
if it satisfies ${K} {K}^{-1}=\mathbb{I}$. Again we stress that
${K}^{}_{\beta, \Phi'} $
is a quantum amplitude computed in the specific spacetime manifold ${ M_D \cup_{B; \hat{K}^{-1}} M_D}$.
\end{widetext}

We now go back to derive
Eqs.(\ref{Z2Dcut}),(\ref{eq:S2S1S1inS4}) and (\ref{eq:S1S2S2inS4}).
For Eq.(\ref{Z2Dcut}), the only path integral 
we need to compute more explicitly is this:
\begin{align}
& Z \bpm \includegraphics[scale=0.3]{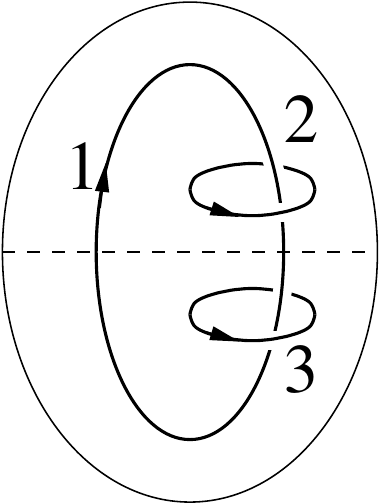} \epm 
= 
 \<0_{D^2_{xt} \times S^1_y}|(W^{S^1_y}_{\si_1})^\dag \hat{\cS} W^{S^1_y}_{\si_2}W^{S^1_y}_{\si_3}|0_{D^2_{xt} \times S^1_y}\> 
\nonumber\\
&= \<0_{D^2_{xt} \times S^1_y}|(W^{S^1_y}_{\si_1})^\dag \hat{\cS} W^{S^1_y}_{\si_4} \cF^{\sigma_4}_{\si_2 \si_3} |0_{D^2_{xt} \times S^1_y}\> 
\nonumber\\
&=
\sum_{\al \si_4} (G^\al_{\si_1})^*\cS_{\al \si_4} \cF^{\si_4}_{\si_2\si_3}=\sum_{\si_4}  \cS_{\bar{\si}_1 \si_4} \cN^{\si_4}_{\si_2\si_3},
\end{align}
where the last equality we use the canonical basis.
Together with the previous data, we can easily derive Eq.(\ref{Z2Dcut}). 

Since it is convenient to express in terms of canonical bases, below for all the derivations, we will implicitly 
project every quantum amplitude into \emph{canonical bases} when we write down its matrix element.

For Eq.(\ref{eq:S2S1S1inS4}), the only path integral 
we need to compute more explicitly is this:
\begin{align}
& Z \bpm \includegraphics[scale=0.35]{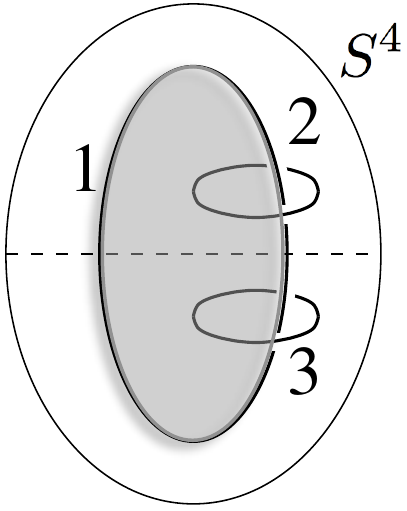} \epm 
= 
 \<0_{D^2_{\varphi w} \times S^2_{\theta\phi}}|(V^{S^2_{\theta\phi}}_{\mu_1})^\dag  W^{S^1_\varphi}_{\sigma_2} W^{S^1_\varphi}_{\sigma_3} | 0_{D^3_{\theta\phi w} \times S^1_{\varphi}} \rangle 
\nonumber\\
&=
 \<0_{D^2_{\varphi w} \times S^2_{\theta\phi}}|(V^{S^2_{\theta\phi}}_{\mu_1})^\dag  W^{S^1_\varphi}_{\sigma_4}  (\cF^{S^1})^{\sigma_4}_{\si_2 \si_3} | 0_{D^3_{\theta\phi w} \times S^1_{\varphi}} \rangle \nonumber\\
&=\sum_{\si_4}  {\tL}^\text{($S^2$,$S^1$)}_{\mu_1\si_4}  (\cF^{S^1})^{\si_4}_{\si_2\si_3},
\end{align}
again we use the canonical basis.
Together with the previous data, we can easily derive Eq.(\ref{eq:S2S1S1inS4}). 
Similarly, we can derive Eq.(\ref{eq:S1S2S2inS4}) using the almost equivalent computation.


Now let us derive Eq.(\ref{eq:Spin[HopfLink]S4}). 
In the first path integral, 
we create a pair of loop $\mu_1$ and anti-loop $\bar{\mu}_1$ excitations and then annihilate them, in terms of
the spacetime picture, 
\be \label{eq:S4T2}
Z \bpm \includegraphics[scale=0.35]{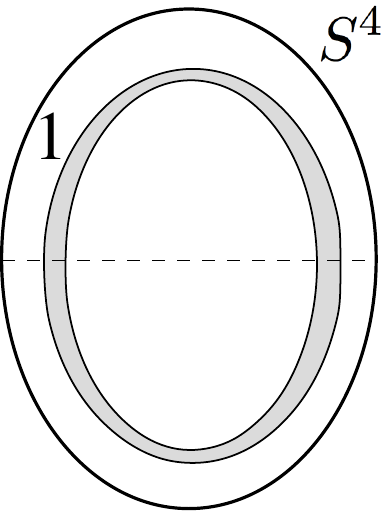} \epm=Z \bpm \includegraphics[scale=0.35]{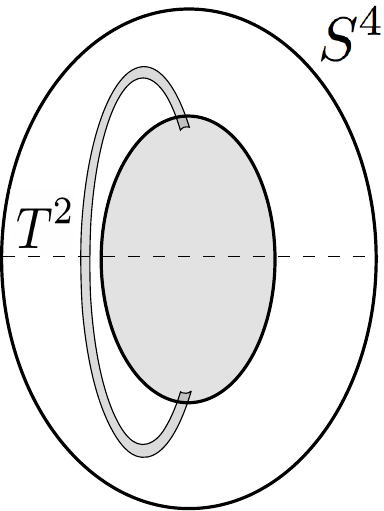} \epm={ \tL^{\text{Tri}}_{0,0,\mu_1}},
\ee
based on the data defined earlier.

In the third path integral 
${\tL^{\text{Tri}}_{{\mu_3}, {\mu_2}, {\mu_1}}}$ of Eq.(\ref{eq:Spin[HopfLink]S4}), there are two descriptions to interpret it in terms of the braiding process in spacetime.
Here is the first description. we create a pair of loop $\mu_1$ and anti-loop $\bar{\mu}_1$ excitations and then
there a pair of  $\mu_2$-$\bar{\mu}_2$ and another pair of $\mu_3$-$\bar{\mu}_3$ are created
while both pairs are thread by $\mu_1$. 
Then the $\mu_1$-$\mu_2$-${\mu_3}$ will do the three-loop braiding process, which gives the most important Berry phase or Berry matrix information into
the path integral. 
After then the pair of  $\mu_2$-$\bar{\mu}_2$ is annihilated and also the pair of $\mu_3$-$\bar{\mu}_3$ is annihilated,
while all the four loops are threaded by $\mu_1$ during the process.
Finally we annihilate the pair of $\mu_1$ and $\bar{\mu}_1$ in the end \cite{Jiang:2014ksa}.
The second description is that we take a Hopf link of $\mu_2$-${\mu_3}$ linking spinning around the loop of $\mu_1$ \cite{Jian:2014vfa, Bi:2014vaa}.
We denote the Hopf link of $\mu_2$-${\mu_3}$ as Hopf$[\mu_3,\mu_2]$, denote its spinning as Spun[Hopf$[\mu_3,\mu_2]]$,
and denote its linking with the third $T^2$-worldsheet of $\mu_1$ as Link[Spun[Hopf$[\mu_3,\mu_2]],\mu_1]$.
Thus we can define ${\tL^{\text{Tri}}_{{\mu_3}, {\mu_2}, {\mu_1}}}  \equiv Z[S^4; \text{Link[Spun[Hopf}[\mu_3,\mu_2]],\mu_1]]$.
From the second description, we immediate see that
${\tL^{\text{Tri}}_{{\mu_3}, {\mu_2}, {\mu_1}}}$ as $Z[S^4; \text{Link[Spun[Hopf}[\mu_3,\mu_2]],\mu_1]]$
are symmetric under exchanging $\mu_2 \leftrightarrow \mu_3$, up to an overall conjugation due to the orientation of quasi-excitations.

We can view the spacetime $S^4$ as a $S^4=\R^4 + \{\infty \}$, the Cartesian coordinate $\R^4$ plus a point at the infinity $ \{\infty \}$.
Similar to the embedding of Ref.\cite{Jian:2014vfa}, we embed the $T^2$-worldsheets ${\mu_1}, {\mu_2}, {\mu_3}$ into the 
$(X_1,X_2,X_3,X_4) \in \R^4$ as follows:
\bea
\left\{
    \begin{array}{l}
   X_1(u, \vec{x})=[r_1(u)+(r_2(u)+r_3(u) \cos x) \cos y] \cos z,\\
   X_2(u, \vec{x})=[r_1(u)+(r_2(u)+r_3(u) \cos x) \cos y] \sin z,\\
   X_3(u, \vec{x})=(r_2(u)+r_3(u) \cos x) \sin y ,\\
   X_4(u, \vec{x})=r_3(u) \sin x,  \end{array}
    \right.\;\;\;\;
\eea
here $\vec{x}\equiv(x,y,z)$.
We choose the $T^2$-worldsheets as follows. \\
\noindent
The $T^2$-worldsheet ${\mu_1}$ is parametrized by some fixed $u_1$ and free coordinates of $(z,x)$ while $y=0$ is fixed.\\
\noindent
The $T^2$-worldsheet ${\mu_2}$ is parametrized by some fixed $u_2$ and free coordinates of $(x,y)$ while $z=0$ is fixed.\\
\noindent
The $T^2$-worldsheet ${\mu_3}$ is parametrized by some fixed $u_3$ and free coordinates of $(y,z)$ while $x=0$ is fixed.\\
We can set the parameters $u_1 > u_2 > u_3$. 
Meanwhile, a $T^3$-surface can be defined as
$\cM^3(u, \vec{x}) \equiv (X_1(u, \vec{x}),X_2(u, \vec{x}),X_3(u, \vec{x}),X_4(u, \vec{x}))$ with a fixed $u$ and free parameters $\vec{x}$.
The $T^3$-surface $\cM^3(u, \vec{x}) \equiv (X_1(u, \vec{x}),X_2(u, \vec{x}),X_3(u, \vec{x}),X_4(u, \vec{x}))$ encloses a 4-dimensional volume.
We define the enclosed 4-dimensional volume as the $\cM^3(u, \vec{x}) \times I^1(s)$ where $I^1(s)$ is the 1-dimensional radius interval along $r_3$,
such that
$I^1(s)=\{ s| s=[0, r_3(u)]\}$, namely $0 \leq s \leq r_3(u)$. Here we can define
$r_3(0)=0$. 
The topology of the enclosed 4-dimensional volume of $\cM^3(u, \vec{x}) \times I^1(s)$
 is of course the $T^3 \times I^1=T^2 \times (S^1 \times I^1) =T^2 \times D^2$.
For a $\cM^3(u_{\text{large}}, \vec{x})\times I^1(s)$ prescribed by a fixed larger $u_{\text{large}}$ and free parameters $\vec{x}$, the $\cM^3(u_{\text{large}}, \vec{x})\times I^1(s)$ must 
enclose the 4-volume spanned by the past history of $\cM^3(u_{\text{small}}, \vec{x})\times I^1(s)$, for any $u_{\text{large}}>u_{\text{small}}$.
Here we set $u_1> u_2> u_3$. And
we also set $r_1(u) > r_2(u) > r_3(u)$ for any given $u$.

One can check that the three $T^2$-worldsheet ${\mu_1},{\mu_2}$ and ${\mu_3}$ indeed have the nontrivial \emph{triple-linking number} \cite{carter2004surfaces}. 
We can design the triple-linking number to be: Tlk$(\mu_2,\mu_1,\mu_3)$=Tlk$(\mu_3,\mu_1,\mu_2)=0$,
Tlk$(\mu_1,\mu_2,\mu_3)=+1$, Tlk$(\mu_3,\mu_2,\mu_1)=-1$, 
Tlk$(\mu_2,\mu_3,\mu_1)=+1$, Tlk$(\mu_1,\mu_3,\mu_2)=-1$. 

Below we will frequently use the surgery trick by cutting out a tubular neighborhood $D^2 \times T^2$ of the $T^2$-worldsheet
and re-gluing this $D^2 \times T^2$ back to its complement $S^4 \- D^2 \times T^2$ via the
modular $\cS^{xyz}$-transformation. The $\cS^{xyz}$-transformation sends
\bea
\bpm x_{\text{out}} \\
y_{\text{out}}  \\
z_{\text{out}} \epm=\bpm 0& 0&1 \\1& 0&0 \\0& 1&0\epm 
\bpm
x_{\text{in}} \\
y_{\text{in}}  \\
z_{\text{in}}\epm.
\eea 
Thus, the $\cS^{xyz}$-identification is 
$(x_{\text{out}},
y_{\text{out}},
z_{\text{out}}) \leftrightarrow 
(z_{\text{in}},
x_{\text{in}},
y_{\text{in}})$.
The $(\cS^{xyz})^{-1}$-identification is 
$(x_{\text{out}},
y_{\text{out}},
z_{\text{out}}) \leftrightarrow 
(y_{\text{in}},
z_{\text{in}},
x_{\text{in}})$.
The surgery on the initial $S^4$ outcomes a new manifold,
\be
(D^2 \times T^2) \cup_{T^3; \cS^{xyz}} (S^4 \- D^2 \times T^2) = S^3 \times S^1 \# S^2 \times S^2.
\ee

In terms of the spacetime path integral picture, use Eqs.(\ref{eq:surger1}) and (\ref{eq:surger2}), we derive:
\begin{widetext}
\bea
&&Z \bpm \includegraphics[scale=0.45]{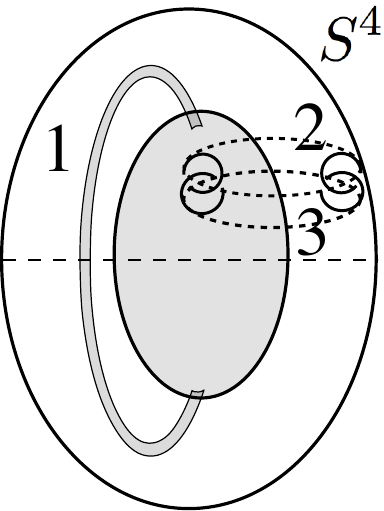} \epm \equiv {\tL^{\text{Tri}}_{{\mu_3}, {\mu_2}, {\mu_1}}}=Z[S^4; \text{Link[Spun[Hopf}[\mu_3,\mu_2]],\mu_1]] \nonumber\\
&&=\sum_{\mu_3'} {\cS^{xyz}_{\mu_3', \mu_3}} \; {Z(S^3 \times S^1 \# S^2 \times S^2; \mu_1, \mu_2 \parallel \mu_3')}  \label{eq:Sxyzsurger1} \\
&&=\sum_{\mu_3', {\Gamma_2}} {\cS^{xyz}_{\mu_3', \mu_3}} (\cF^{T^2})_{\mu_2 \mu_3'}^{\Gamma_2}  \; {Z(S^3 \times S^1 \# S^2 \times S^2; \mu_1, \Gamma_2)} \label{eq:Sxyzsurger2}\\
&&=\sum_{\mu_3', {\Gamma_2}, {\Gamma_2'}} {\cS^{xyz}_{\mu_3', \mu_3}} (\cF^{T^2})_{\mu_2 \mu_3'}^{\Gamma_2} {(\cS^{xyz})^{-1}_{\Gamma_2', \Gamma_2}} \; {Z(S^4; \mu_1, \Gamma_2')} \label{eq:Sxyzsurger3}\\
&&=\sum_{\mu_3', {\Gamma_2}, {\Gamma_2'}, {\Gamma_2''}} {\cS^{xyz}_{\mu_3', \mu_3}} (\cF^{T^2})_{\mu_2 \mu_3'}^{\Gamma_2} {(\cS^{xyz})^{-1}_{\Gamma_2', \Gamma_2}} 
{(\cS^{xyz})^{-1}_{\Gamma_2'', \Gamma_2'}} \; {Z(S^3 \times S^1 \# S^2 \times S^2;  \mu_1, \Gamma_2'')} \label{eq:Sxyzsurger4}\\
&&=\sum_{\mu_3', {\Gamma_2}, {\Gamma_2'}, {\Gamma_2''}, {\eta_2}} {\cS^{xyz}_{\mu_3', \mu_3}} (\cF^{T^2})_{\mu_2 \mu_3'}^{\Gamma_2} {(\cS^{xyz})^{-1}_{\Gamma_2', \Gamma_2}} 
{(\cS^{xyz})^{-1}_{\Gamma_2'', \Gamma_2'}}  (\cF^{T^2})_{{\mu_1} {\Gamma_2''}}^{\eta_2}\; {Z(S^3 \times S^1 \# S^2 \times S^2;  \eta_2)} \label{eq:Sxyzsurger5}\\
&&=\sum_{\mu_3', {\Gamma_2}, {\Gamma_2'}, {\Gamma_2''}, {\eta_2}, {\eta_2'}} {\cS^{xyz}_{\mu_3', \mu_3}} (\cF^{T^2})_{\mu_2 \mu_3'}^{\Gamma_2} {(\cS^{xyz})^{-1}_{\Gamma_2', \Gamma_2}} 
{(\cS^{xyz})^{-1}_{\Gamma_2'', \Gamma_2'}}  (\cF^{T^2})_{{\mu_1} {\Gamma_2''}}^{\eta_2}  {\cS^{xyz}_{\eta_2', \eta_2}}  \; {\tL^{\text{Tri}}_{0, 0, {\eta_2'}}} \label{eq:Sxyzsurger6}. 
\eea
As usual, the repeated indices are summed over. 
With the trick of $\cS^{xyz}$-transformation in mind, here is the step-by-step sequence of surgeries we perform.  
\end{widetext}

\noindent
Step 1: We cut out the tubular neighborhood $D^2 \times T^2$ of the $T^2$-worldsheet of $\mu_3$ and re-glue this $D^2 \times T^2$ back to its complement $S^4 \- D^2 \times T^2$ via the
modular $(\cS^{xyz})^{-1}$-transformation. The $D^2 \times T^2$ neighborhood of $\mu_3$-worldsheet can be viewed as the 4-volume 
$\cM^3(u_{3}, \vec{x})\times I^1(s)$, which encloses neither $\mu_1$-worldsheet nor $\mu_2$-worldsheet. 
The $(\cS^{xyz})^{-1}$-transformation sends $(y_{\text{in}},
z_{\text{in}})$ of $\mu_3$ to
$(x_{\text{out}},
y_{\text{out}})$ of $\mu_2$. The gluing however introduces the summing-over new coordinate $\mu_3'$, 
based on Eq.(\ref{eq:surger1}). Thus Step 1 obtains Eq.(\ref{eq:Sxyzsurger1}).\\

In Step 1, as Eq.(\ref{eq:Sxyzsurger1}) and thereafter, we write down ${\cS^{xyz}_{\mu_3', \mu_3}}$ matrix.
Based on Eq.(\ref{eq:projectK}), 
we stress that the ${\cS^{xyz}_{\mu_3', \mu_3}}$ is projected to the $|0_{D^2 \times T^2} \rangle$-states with operator-insertions for both bra and ket states.\\ 
\bea
\label{eq:projectS}
&&{\cS^{xyz}_{\mu_3', \mu_3}} \equiv \< {\mu_3'}_{D^2 \times T^2}|  \hat{\cS}^{xyz} | {\mu_3}_{D^2 \times T^2} \rangle_{ {D^2 \times T^2} \cup_{T^3; \hat{S}^{xyz}} {D^2 \times T^2}} \nonumber \\
&&=\< 0_{D^2_{xw} \times T^2_{yz}} | V^{T^2_{yz} \dagger}_{\mu_3'}  \hat{\cS}^{xyz}  V^{T^2_{yz}}_{\mu_3}  | 0_{D^2_{xw} \times T^2_{yz}} \rangle_{S^3 \times S^1} 
\eea
Here we use the surgery fact 
\be
{ {D^2 \times T^2} \cup_{T^3; \hat{S}^{xyz}} {D^2 \times T^2}}={S^3 \times S^1}.
\ee
So our ${\cS^{xyz}_{\mu_3', \mu_3}}$ is defined as a quantum amplitude in ${S^3 \times S^1}$. Two $T^2$-worldsheets
${\mu_3'}$ and ${\mu_3}$ now become a pair of Hopf link resides in $S^3$ part of ${S^3 \times S^1}$, while
share the same $S^1$ circle in the $S^1$ part of ${S^3 \times S^1}$.
We can view the shared $S^1$ circle as the spinning circle of the spun surgery construction 
on the Hopf link in $D^3$, the spun-topology would be
$D^3 \times S^1$, then we glue this $D^3 \times S^1$ contains $\text{Spun[Hopf}[{\mu_3'},{\mu_3}]]$ to another $D^3 \times S^1$, so 
we have $D^3 \times S^1 \cup_{S^2 \times S^1} D^3 \times S^1= S^3 \times S^1$ as an overall new spacetime topology.
Hence we also denote 
\be
{\cS^{xyz}_{\mu_3', \mu_3}}=Z[{S^3 \times S^1}; \text{Spun[Hopf}[{\mu_3'},{\mu_3}]]]. 
\ee

\noindent
Step 2: The earlier surgery now makes the inner $\mu_3'$-worldsheet \emph{parallels} to the outer $\mu_2$-worldsheet,
since they share the same coordinates $(x_{\text{out}}, y_{\text{out}})=(y_{\text{in}},z_{\text{in}})$. We denote their parallel topology as $\mu_2 \parallel \mu_3'$.
So we can fuse 
the $\mu_2$-worldsheet and $\mu_3'$-worldsheet via the fusion algebra, namely
$V^{T^2_{x_{\text{out}},
y_{\text{out}}}}_{\mu_2} V_{\mu_3'}^{T^2_{x_{\text{out}},
y_{\text{out}}}}=  (\cF^{T^2})_{\mu_2 \mu_3'}^{\Gamma_2} V_{\Gamma_2}^{T^2_{x_{\text{out}},
y_{\text{out}}}}$. Thus Step 2 obtains Eq.(\ref{eq:Sxyzsurger2}). \\

\noindent
Step 3: We cut out the tubular neighborhood $D^2 \times T^2$ of the $T^2$-worldsheet of ${\Gamma_2}$ and re-glue this $D^2 \times T^2$ back to its complement $S^4 \- D^2 \times T^2$ via the
modular $\cS^{xyz}$-transformation. The $D^2 \times T^2$ neighborhood of ${\Gamma_2}$-worldsheet can be viewed as the 4-volume 
$\cM^3(u_{2}, \vec{x})\times I^1(s)$ in the new manifold $S^3 \times S^1 \# S^2 \times S^2$, which encloses no worldsheet inside. 
After the surgery, 
the $\cS^{xyz}$-transformation sends the redefined $(x_{\text{in}},
y_{\text{in}})$ of $\Gamma_2$ back to
$(y_{\text{out}},
z_{\text{out}})$ of $\Gamma_2'$. The gluing however introduces the summing-over new coordinate $\Gamma_2'$, 
based on Eq.(\ref{eq:surger1}). 
We also transform $S^3 \times S^1 \# S^2 \times S^2$ back to $S^4$ again. 
Thus Step 3 obtains Eq.(\ref{eq:Sxyzsurger3}).\\

\noindent
Step 4: We cut out the tubular neighborhood $D^2 \times T^2$ of the $T^2$-worldsheet of ${\Gamma_2'}$ and re-glue this $D^2 \times T^2$ back to its complement $S^4 \- D^2 \times T^2$ via the
modular $\cS^{xyz}$-transformation. The $D^2 \times T^2$ neighborhood of ${\Gamma_2}'$-worldsheet viewed as the 4-volume 
in the manifold $S^4$ encloses no worldsheet inside. 
After the surgery, 
the $\cS^{xyz}$-transformation sends the $(x_{\text{in}},
y_{\text{in}})$ of $\Gamma_2'$ to
$(z_{\text{out}},
x_{\text{out}})$ of $\mu_1$. The gluing however introduces the summing-over new coordinate $\Gamma_2''$, 
based on Eq.(\ref{eq:surger1}). 
We also transform $S^4$ to  $S^3 \times S^1 \# S^2 \times S^2$ again. 
Thus Step 4 obtains Eq.(\ref{eq:Sxyzsurger4}).\\

\noindent
Step 5: The earlier surgery now makes the inner $\Gamma_2''$-worldsheet \emph{parallels} to the outer $\mu_1$-worldsheet,
since they share the same coordinates $(z_{\text{out}},
x_{\text{out}})=(x_{\text{in}},
y_{\text{in}})$. We denote their parallel topology as $\mu_1 \parallel \Gamma_2''$.
We now fuse 
the $\mu_1$-worldsheet and $\Gamma_2''$-worldsheet via the fusion algebra, namely
$V_{\mu_1}^{T^2_{z_{\text{out}},
x_{\text{out}}}} V_{\Gamma_2''}^{T^2_{z_{\text{out}},
x_{\text{out}}}}=  (\cF^{T^2})_{{\mu_1} {\Gamma_2''}}^{\eta_2} V_{\eta_2}^{T^2_{z_{\text{out}},
x_{\text{out}}}}$. Thus Step 5 obtains Eq.(\ref{eq:Sxyzsurger5}). \\ 

\noindent
Step 6: We should do the inverse transformation to get back to the $S^4$ manifold. 
Thus we cut out the tubular neighborhood $D^2 \times T^2$ of the $T^2$-worldsheet of ${\eta_2}$ and re-glue this $D^2 \times T^2$ back to its complement via the
modular $(\cS^{xyz})^{-1}$-transformation. We relate the original path integral to the final one ${Z(S^4;  \eta_2')}={\tL^{\text{Tri}}_{0, 0, {\eta_2'}}}$.
  Thus Step 6 obtains Eq.(\ref{eq:Sxyzsurger6}). 

\begin{widetext}
Similarly, in the fourth path integral of Eq.(\ref{eq:Spin[HopfLink]S4}), we derive
\bea \label{eq:SxyzsurgerDown}
&&Z \bpm \includegraphics[scale=0.45]{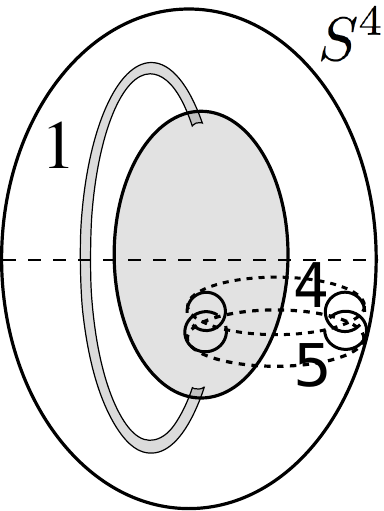} 
\epm \equiv {\tL^{\text{Tri}}_{{\mu_5}, {\mu_4}, {\mu_1}}}=Z[S^4; \text{Link[Spun[Hopf}[\mu_5,\mu_4]],\mu_1]] \nonumber\\
&&=\sum_{\mu_5', {\Gamma_4}, {\Gamma_4'}, {\Gamma_4''}, {\eta_4}, {\eta_4'}} {\cS^{xyz}_{\mu_5', \mu_5}} (\cF^{T^2})_{\mu_4 \mu_5'}^{\Gamma_4} {(\cS^{xyz})^{-1}_{\Gamma_4', \Gamma_4}} 
{(\cS^{xyz})^{-1}_{\Gamma_4'', \Gamma_4'}}  (\cF^{T^2})_{{\mu_1} {\Gamma_4''}}^{\eta_4}  {\cS^{xyz}_{\eta_4', \eta_4}}  \; {\tL^{\text{Tri}}_{0, 0, {\eta_4'}}}. 
\eea

In the second path integral of Eq.(\ref{eq:Spin[HopfLink]S4}), 
we have the Hopf link of $\text{Hopf}[\mu_3,\mu_2]$ and the Hopf link of $\text{Hopf}[\mu_5,\mu_4]$. 
In the spacetime picture, all $\mu_2, \mu_3, \mu_4, \mu_5$ are $T^2$-worldsheets under the spun surgery construction. 
We can locate the the spun object named $\text{Spun[Hopf}[\mu_3,\mu_2],\text{Hopf}[\mu_5,\mu_4]]$ inside a $D^3 \times S^1$, while this
$D^3 \times S^1$ is glued with a $S^2 \times D^2$ to a $S^4$. Here the $S^2 \times D^2$ contains a $T^2$-worldsheet $\mu_1$.
We can view the $T^2$-worldsheet $\mu_1$ contains a $S^2$-sphere of the $S^2 \times D^2$ but attached an extra handle.
We derive: 
\bea
&&Z \bpm \includegraphics[scale=0.45]{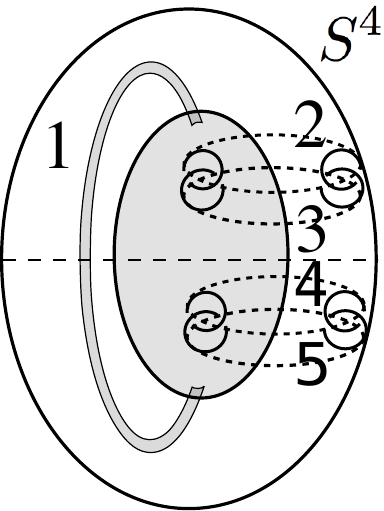} 
\epm \equiv Z[S^4; \text{Link[Spun[Hopf}[\mu_3,\mu_2],\text{Hopf}[\mu_5,\mu_4]]],\mu_1]] \nonumber\\
&&=
\sum_{\mu_3', {\Gamma_2}, {\Gamma_2'}}
\sum_{\mu_5', {\Gamma_4}, {\Gamma_4'}}
{\cS^{xyz}_{\mu_3', \mu_3}} (\cF^{T^2})_{\mu_2 \mu_3'}^{\Gamma_2} {(\cS^{xyz})^{-1}_{\Gamma_2', \Gamma_2}} 
{\cS^{xyz}_{\mu_5', \mu_5}} (\cF^{T^2})_{\mu_4 \mu_5'}^{\Gamma_4} {(\cS^{xyz})^{-1}_{\Gamma_4', \Gamma_4}}
Z[S^4; \text{Spun}[\Gamma_2',\Gamma_4'],\mu_1] \label{eq:DoubleHopf1}  \\
&&=
\sum_{\mu_3', {\Gamma_2}, {\Gamma_2'}}
\sum_{\mu_5', {\Gamma_4}, {\Gamma_4'}} \sum_{\Gamma}
{\cS^{xyz}_{\mu_3', \mu_3}} (\cF^{T^2})_{\mu_2 \mu_3'}^{\Gamma_2} {(\cS^{xyz})^{-1}_{\Gamma_2', \Gamma_2}} 
{\cS^{xyz}_{\mu_5', \mu_5}} (\cF^{T^2})_{\mu_4 \mu_5'}^{\Gamma_4} {(\cS^{xyz})^{-1}_{\Gamma_4', \Gamma_4}}
(\cF^{T^2})_{\Gamma_2',\Gamma_4'}^{\Gamma} Z[S^4; \Gamma,\mu_1] \label{eq:DoubleHopf2} \;\;\;  \\
&&=\sum_{\overset{\mu_3', {\Gamma_2}, {\Gamma_2'}}{\mu_5', {\Gamma_4}, {\Gamma_4'}}}
\sum_{\Gamma, \Gamma', \Gamma_1, \Gamma_1'}
{\cS^{xyz}_{\mu_3', \mu_3}} (\cF^{T^2})_{\mu_2 \mu_3'}^{\Gamma_2} {(\cS^{xyz})^{-1}_{\Gamma_2', \Gamma_2}} 
{\cS^{xyz}_{\mu_5', \mu_5}} (\cF^{T^2})_{\mu_4 \mu_5'}^{\Gamma_4} {(\cS^{xyz})^{-1}_{\Gamma_4', \Gamma_4}}
(\cF^{T^2})_{\Gamma_2',\Gamma_4'}^{\Gamma}
{(\cS^{xyz})^{-1}_{\Gamma', \Gamma}}  (\cF^{T^2})_{{\mu_1} {\Gamma'}}^{\Gamma_1}  {\cS^{xyz}_{\Gamma_1', \Gamma_1}}  \; {\tL^{\text{Tri}}_{0, 0, {\Gamma_1'}}}. \label{eq:DoubleHopf3}\;\;\;\;\;\;\;\;\;
\eea
Here we do the Step 1, Step 2 and Step 3 surgeries on $\text{Spun[Hopf}[\mu_3,\mu_2]]$ first, then do the same 3-step surgeries on $\text{Spun[Hopf}[\mu_5,\mu_4]]$ later,
then we obtain Eq.(\ref{eq:DoubleHopf1}).
While in Eq.(\ref{eq:DoubleHopf1}), the new $T^2$-worldsheets ${\Gamma_2'}$ and ${\Gamma_4'}$ have no triple-linking with the worldsheet $\mu_1$. 
Here ${\Gamma_2'}$ and ${\Gamma_4'}$ are arranged in the $D^3 \times S^1$ part of the $S^4$ manifold, while $\mu_1$ is in the $S^2 \times D^2$ part of the $S^4$ manifold.
Indeed, ${\Gamma_2'}$ and ${\Gamma_4'}$ can be fused together in parallel to a new $T^2$-worldsheet ${\Gamma}$ via the fusion algebra $(\cF^{T^2})_{\Gamma_2',\Gamma_4'}^\Gamma$, so we obtain Eq.(\ref{eq:DoubleHopf2}).
Then we apply the Step 4, Step 5 and Step 6 surgeries on the $T^2$-worldsheets $\Gamma$ and $\mu_1$ of $Z[S^4; \text{Spun}[\Gamma],\mu_1]= Z[S^4; \Gamma,\mu_1]$ in Eq.(\ref{eq:DoubleHopf2}),
we obtain the final form Eq.(\ref{eq:DoubleHopf3}).

Use Eqs.(\ref{eq:S4T2}),(\ref{eq:Sxyzsurger6}),(\ref{eq:SxyzsurgerDown}) and (\ref{eq:DoubleHopf3}), and plug them into the path integral surgery relations, we derive a new quantum surgery formula (namely Eq.(\ref{eq:S2S1S1inS4}) in the main text):
\bea
\label{}
&&\ \ \ \
  Z \bpm \includegraphics[scale=0.35]{3+1D_Triple_link_T2_1.pdf} \epm 
 Z \bpm \includegraphics[scale=0.35]{3+1D_Triple_link_double_SpinHopfLink_S2_double_12345.pdf} \epm 
=
 Z \bpm \includegraphics[scale=0.35]{3+1D_Triple_link_double_SpinHopfLink_S2_Up_123.pdf} \epm 
 Z \bpm \includegraphics[scale=0.35]{3+1D_Triple_link_double_SpinHopfLink_S2_Down_145.pdf} \epm \nonumber \\
&& \Rightarrow
 {{\tL^{\text{Tri}}_{0, 0, {\mu_1}}} \cdot
 \sum_{\Gamma, \Gamma', \Gamma_1, \Gamma_1'}
 (\cF^{T^2})_{\Gamma_2',\Gamma_4'}^{\Gamma}
{(\cS^{xyz})^{-1}_{\Gamma', \Gamma}}  (\cF^{T^2})_{{\mu_1} {\Gamma'}}^{\Gamma_1}  {\cS^{xyz}_{\Gamma_1', \Gamma_1}}  \; {\tL^{\text{Tri}}_{0, 0, {\Gamma_1'}}} } \nonumber
\\
&&
{=
{\sum_{{\Gamma_2''}, {\eta_2}, {\eta_2'}} 
{(\cS^{xyz})^{-1}_{\Gamma_2'', \Gamma_2'}}  (\cF^{T^2})_{{\mu_1} {\Gamma_2''}}^{\eta_2}  {\cS^{xyz}_{\eta_2', \eta_2}}  \; {\tL^{\text{Tri}}_{0, 0, {\eta_2'}}} 
} 
\cdot
{\sum_{{\Gamma_4''}, {\eta_4}, {\eta_4'}}  
{(\cS^{xyz})^{-1}_{\Gamma_4'', \Gamma_4'}}  (\cF^{T^2})_{{\mu_1} {\Gamma_4''}}^{\eta_4}  {\cS^{xyz}_{\eta_4', \eta_4}}  \; {\tL^{\text{Tri}}_{0, 0, {\eta_4'}}}
 }}, 
\eea
here only ${\mu_1},{\Gamma_2'},{\Gamma_4'}$ are the fixed indices, other indices are summed over.
\end{widetext}

Lastly we provide more explicit calculations of Eq.(\ref{eq:FS1FT2}), the constraint between the fusion data itself.
First, we recall that
\bea
&& W_{\sigma_1}^{S^1_y}  W_{\sigma_2}^{S^1_y} = (\cF^{S^1})_{{\sigma_1}{\sigma_2}}^{\sigma_3} W_{\sigma_3}^{S^1_y},\\
&&  V_{\mu_1}^{T^2_{yz}} V_{\mu_2}^{T^2_{yz}}=  (\cF^{T^2})_{{\mu_1}{\mu_2}}^{\mu_3} V_{\mu_3}^{T^2_{yz}},\\
&& W_{\sigma_1}^{S^1_y} V_{\mu_2}^{T^2_{yz}}=(\cF^{T^2})_{{\sigma_1}{\mu_2}}^{\mu_3} V_{\mu_3}^{T^2_{yz}}.
\eea
Of course, the fusion algebra is symmetric respect to exchanging the lower indices, $(\cF^{T^2})_{{\sigma_1}{\mu_2}}^{\mu_3}=(\cF^{T^2})_{{\mu_2}{\sigma_1}}^{\mu_3}$.
We can regard the fusion algebra $(\cF^{S^1})_{{\sigma_1}{\sigma_2}}^{\sigma_3}$ and $(\cF^{T^2})_{{\sigma_1}{\mu_2}}^{\mu_3}$ with worldlines
as a part of a larger algebra of the fusion algebra of worldsheets $(\cF^{T^2})_{{\mu_1}{\mu_2}}^{\mu_3}$.
We compute the state $W_{\sigma_1}^{S^1_y}  W_{\sigma_2}^{S^1_y} V_{\mu_2}^{T^2_{yz}} | 0_{D^2_{wx} \times T^2_{yz}} \rangle$
by fusing two $W^{S^1}_{\si}$ operators and one $V^{T^2}_{\mu}$ operator in different orders.

On one hand, we can fuse two worldlines first, then fuse with the worldsheet,
\bea
&& W_{\sigma_1}^{S^1_y}  W_{\sigma_2}^{S^1_y} V_{\mu_2}^{T^2_{yz}} | 0_{D^2_{wx} \times T^2_{yz}} \rangle \nonumber \\
&&=\sum_{{\sigma_3}} (\cF^{S^1})_{{\sigma_1}{\sigma_2}}^{\sigma_3} W_{\sigma_3}^{S^1_y} V_{\mu_2}^{T^2_{yz}} | 0_{D^2_{wx} \times T^2_{yz}} \rangle \nonumber\\
&&=\sum_{{\sigma_3},{\mu_3}} (\cF^{S^1})_{{\sigma_1}{\sigma_2}}^{\sigma_3}   (\cF^{T^2})_{{\sigma_3}{\mu_2}}^{\mu_3}  V_{\mu_3}^{T^2_{yz}} | 0_{D^2_{wx} \times T^2_{yz}} \rangle. \label{eq:WWV1}
\eea
On the other hand, we can fuse a worldline with the worldsheet first, then fuse with another worldline,
\bea
&& W_{\sigma_1}^{S^1_y}  W_{\sigma_2}^{S^1_y} V_{\mu_2}^{T^2_{yz}} | 0_{D^2_{wx} \times T^2_{yz}} \rangle \nonumber\\
&&=\sum_{{\mu_1}} W_{\sigma_1}^{S^1_y} (\cF^{T^2})_{{\sigma_2}{\mu_2}}^{\mu_1}  V_{\mu_1}^{T^2_{yz}}  | 0_{D^2_{wx} \times T^2_{yz}} \rangle \nonumber\\
&&=\sum_{{\mu_1},{\mu_3}} (\cF^{T^2})_{{\sigma_2}{\mu_2}}^{\mu_1} (\cF^{T^2})_{{\sigma_1}{\mu_1}}^{\mu_3}
V_{\mu_3}^{T^2_{yz}}  | 0_{D^2_{wx} \times T^2_{yz}} \rangle \label{eq:WWV2}
\eea
Therefore, by comparing Eqs.(\ref{eq:WWV1}) and (\ref{eq:WWV2}),
we derive a consistency condition for fusion algebra:
\bea
\sum_{\sigma_3} (\cF^{S^1})_{{\sigma_1}{\sigma_2}}^{\sigma_3}   (\cF^{T^2})_{{\sigma_3}{\mu_2}}^{\mu_3}
=\sum_{\mu_1} (\cF^{T^2})_{{\sigma_2}{\mu_2}}^{\mu_1} (\cF^{T^2})_{{\sigma_1}{\mu_1}}^{\mu_3}. \;\;\;\;\; 
\eea

\begin{widetext}
\end{widetext}




\begin{thebibliography}{48}
\bibitem [{Rep()}]{Representation}%
  \BibitemOpen
  \href@noop {} {\emph {\bibinfo {title} {\emph{If there is a gauge theory
  description, then the quasi-excitation type of particle ${\si}$ and string
  $\mu$ would be labeled by the representation for gauge charge and by the
  conjugacy class of the gauge group for gauge flux.}}}}\BibitemShut {Stop}%
\bibitem [{FM()}]{FM}%
  \BibitemOpen
  \href@noop {} {\emph {\bibinfo {title} {\emph{To be even more specific, we
  can also specify the whole open-manifold topology ${M \times V}$
  corresponding to the ground state sector $|0_{{M \times V}} \>$, namely we
  can rewrite the data in new notations to introduce the refined data:
  $(F^{S^1}) \to (F^{D^3 \times S^1})$, $(F^{T^2}) \to (F^{D^2 \times T^2})$
  and $(F^{S^2}) \to (F^{D^2 \times S^2})$. However, because $(F^{M})$ are the
  fusion data in the local neighborhood around the worldline/worldsheet
  operators with topology $M$, physically it does not encode the information
  from the remained product space ${M \times V}$. Namely, at least for the most
  common and known theory that we can regularize on the lattice, we understand
  that $(F^{M})=(F^{M \times V_1})=(F^{M \times V_2})=\dots$ for any topology
  $V_1, V_2, \dots$. So here we make a physical assumption that $(F^{S^1}) =
  (F^{D^3 \times S^1})$, $(F^{T^2}) = (F^{D^2 \times T^2})$ and $(F^{S^2}) =
  (F^{D^2 \times S^2})$ }}}}\BibitemShut {NoStop}%
\end{thebibliography}

\end{document}